\documentclass[useAMS,usenatbib,usegraphicx]{mn2e}

\usepackage{graphicx}
\usepackage{amsmath, amssymb}

\begin{document}
\title[Global lopsided instability]{Global lopsided instability in a purely stellar galactic disc}
\author[Saha, Combes  \& Jog]{Kanak Saha,$^1$\thanks{E-mail:
kanak@physics.iisc.ernet.in} Francoise Combes,$^2$ Chanda J. Jog$^1$\\
$^1$Department of Physics, Indian Institute of Science, Bangalore 560012, India\\
$^2$Observatoire de Paris, LERMA, 61 Av.\ de l'Observatoire, 75014, Paris,  France}
\maketitle
\begin{abstract}
It is shown that pure exponential discs in spiral galaxies are capable of supporting slowly varying discrete global lopsided modes, 
which can explain the observed features of lopsidedness in the stellar discs. Using linearized fluid dynamical equations with the 
softened self-gravity and pressure of the perturbation as the collective effect, 
we derive self-consistently a quadratic eigenvalue equation 
for the lopsided perturbation in the galactic disc. On solving this, we find that 
the ground-state mode shows the observed characteristics of the lopsidedness in a galactic disc, namely the fractional Fourier 
amplitude A$_1$ increases smoothly with the radius. These lopsided patterns precess in the disc with a very slow
 pattern speed with no preferred sense of precession. We show that the lopsided modes in the stellar disc are long-lived
 because of a substantial reduction ($\sim$ a factor of 10 compared to the local free precession rate) in the differential precession. 
The numerical solution of the equations shows that the ground-state lopsided modes are either very slowly precessing
 stationary normal mode oscillations of the disc or growing modes with a slow growth rate depending on the relative
 importance of the collective effect of the self-gravity. 
N-body simulations are performed to test the spontaneous growth of lopsidedness
in a pure stellar disc. Both approaches are then compared and interpreted in terms
of long-lived global $m=1$ instabilities, with almost zero pattern speed.
\end{abstract}

\section{Introduction}
There is growing observational evidence that large-scale lopsided asymmetry is a common phenomenon in
 spiral galaxies, as seen in the stellar distribution (e.g., Block et al. 1994, Rix \& Zaritsky 1995, Bournaud et al. 2005) 
as well as in the distribution of the atomic hydrogen gas  (e.g., Baldwin et al. 1980, hence BLS, Richter 
\& Sancisi 1994, Haynes et al. 1998). These observations show that about 30\% of the field spirals show significant disc lopsidedness.

Various theoretical explanations have been put forward towards explaining the physical origin of the lopsided 
mass distribution in spiral galaxies. Some of these are: the disc response to a distorted halo (Jog 1997, 1999), 
satellite infall onto a galaxy (Zaritsky \& Rix 1997), and tidal interaction and asymmetric gas accretion onto a galactic disc (Bournaud et al. 2005). 
Cooperation of orbital streams to the lopsided pattern was used to drive the 
lopsided instabilities in disc galaxies (Earn \& Lynden-Bell 1996).  
Lovelace et al. (1999) found strongly unstable eccentric motions within one disc scalelength, and the model 
with an off-centred disc w.r.t. the halo by Levine \& Sparke (1998) also shows lopsidedness in the inner 
regions- but both these do not give lopsidedness in the outer regions where it is seen in galaxies.  
So far some theoretical studies have shown that lopsided instabilities exist only in counter-rotating stellar 
discs in which the fraction of retrograde stars is large (Hozumi \& Fujiwara 1989; Sellwood \& Valluri 1997). 
Since counter rotation is a rarely seen phenomenon in stellar discs (Kuijken, Fisher, \& Merrifield 1996; 
Kannappan \& Fabricant 2001) it leaves a room for the search of lopsided instability in normal differentially rotating galactic discs.
A kinematic origin for lopsidedness  in stellar disc indicates a short winding-up time scale which is less than a 
Gyr (BLS). N-body simulations show the disc lopsidedness to be long-lived to $\sim$ 3-4 Gyr, however the physical 
reason for this is not understood. It has been shown that in field galaxies, the disc  lopsidedness is uncorrelated to
 the strength of tidal interaction (Bournaud et al. 2005), or the presence of nearby companions (Wilcots \& 
Prescott 2005). The latter point is exemplified by the case of M101 which is an isolated galaxy and yet is strongly lopsided. 
Thus, an internal origin for the generation of the disc lopsidedness is worth investigating. Because it has the potential 
to provide answer to the question: why are so many isolated galaxies lopsided? This provides the motivation for our present work.

The two main observed characteristics of the disc lopsidedness are that: first, A$_1$, the fractional Fourier amplitude
 of lopsidedness increases smoothly with radius as seen in Rix \& Zaritsky (1995), Bournaud et al. (2005), and 
measured between 1.5-2.5 disc scalelengths. Second, the phase of lopsidedness is nearly constant with radius
(Rix \& Zaritsky 1995, Angiras et al. 2006). The latter indicates that the disc lopsidedness is a global feature (e.g., Jog 1997).

Guided by these observed features, in this paper, we study self-consistently the behaviour of the global lopsided 
perturbation in a purely exponential galactic disc using the fluid dynamical approach.  The lopsidedness emerges
 as a global instability arising due to the collective effect of the self-gravity of the perturbation in the disc. 
The resulting pattern speed is almost constant with radius and thus avoids the winding up problem due to the 
differential precession. 
In our picture, the lopsidedness manifests itself as a classic case of the negative damping phenomenon in a 
self-gravitating system. 
Note that the modes with $m=1$ symmetry have in general no inner Lindblad resonance in the disc. So 
refraction and dissipation of acoustic waves is less of a problem (Block et al. 1994).
Unlike the usual $m=2$ spiral pattern, there exists no definite pattern speed measurement for the $m=1$ 
lopsided mode in the galactic disc.

For simplicity and to separate the various different dynamical mechanisms, 
we treat the lopsidedness in a purely exponential stellar galactic disc 
and disregard the effect of halo.  We also postpone the treatment of a dissipative
component.
In a future paper, we will include the effect of halo and treat a 
similar global calculation  that is applicable to the atomic hydrogen gas 
located at larger radii.
We tackle the existence of global $m=1$ mode both from analytical development,
and through numerical simulations.
                  
We organise the paper in the following way: In section 2 we formulate the dynamical 
equations governing the lopsided mode in the disc and provide a numerical scheme for 
solving the matrix eigenvalue equation. The results from this linear approach are 
described in section 3. Section 4 describes the N-body simulations of a pure
stellar disc, and the study of the $m=1$ modes. The comparison between the two
approaches are compared and interpreted in section 5. Section 6 
summarizes our conclusion.

\section{Dynamics of global lopsided mode}

We formulate here the dynamical equations governing the global lopsided mode in a thin, initially axisymmetric 
self-gravitating disc having an exponential distribution of surface density  ($\Sigma^{0}$). By assuming hydrostatic
 equilibrium along the vertical direction, we ignore the vertical structure of the disc by integrating all physical quantities 
over z. The disc is differentially rotating with angular speed $\Omega(R)$ so that the unperturbed velocity field in the
 plane is given by $\vec{v} = (0,R\Omega)$.
We write the velocity field, surface density, potential and the pressure of the perturbed disc as :
 $v_{R}=v_{R}^{\prime}$, $v_{\varphi}= R\Omega + v_{\varphi}^{\prime}$, $\Sigma = \Sigma^{0} + \Sigma^{\prime}$, $\Phi = \Phi^{0} + \Phi^{\prime}$ and $P = P^{0} + P^{\prime}$.  

\noindent Then the linearized Euler and continuity equations in the cylindrical coordinate system ($R,\varphi$) can be written as:
\begin{align*}
&\frac{Dv_{R}^{\prime}}{Dt} - 2\Omega v_{\varphi}^{\prime} \:=\: -\frac{\partial{\Phi^{\prime}}}{\partial R} -\frac{1}{\Sigma^{0}}\frac{\partial P_{R}^{\prime}}{\partial R} + \frac{P_{\varphi}^{\prime} - P_{R}^{\prime}}{R\Sigma^{0}}\: \: \: \: \: \: \: \: \: \: \: \: \: \: \: \: (1) \\
&\frac{Dv_{\varphi}^{\prime}}{Dt} + \frac{\kappa^2}{2\Omega} v_{R}^{\prime} \:=\: -{\frac{1}{R}}\frac{\partial{\Phi^{\prime}}}{\partial \varphi} -{\frac{1}{R\Sigma^{0}}}\frac{\partial P_{\varphi}^{\prime}}{\partial \varphi}\: \: \: \: \: \: \: \: \: \: \: \: \: \: \: \: \: \:\: \: \: \: \: \: \: \: \: \: \: \: \: \: (2)\\
&\frac{D\Sigma^{\prime}}{Dt} + {\frac{1}{R}}\frac{\partial(R\Sigma^{0}v_{R}^{\prime})}{\partial R} + {\frac{\Sigma^{0}}{R}}\frac{\partial{v_{\varphi}^{\prime}}}{\partial \varphi} \:=\: 0 \: \: \: \: \: \: \: \: \: \: \: \: \: \: \: \: \: \: \: \: \: \: \: \: \: \: \: \: \: \: \: \: \: \: \: (3)\\
\end{align*}
\noindent In the above equations ${D}/{Dt}\equiv {\partial}/{\partial t} + \Omega {\partial}/{\partial \varphi}$ and  
$\kappa$ is the epicyclic frequency in the disc.
\noindent We consider two separate cases namely a cold disc and a hot disc throughout this paper. In the case of a cold disc, the equlibrium state is defined by a balance between the differential rotation and the softened gravity. Whereas in the equilibrium state, a hot disc is supported by the differential rotation and pressure against the softened gravity. 
\noindent Since we are primarily interested in the stability properties of the global lopsided modes($m=1$) in the disc,
 we consider all the perturbed variables as $X^{\prime}(R,\varphi,t) = X^{\prime}(R)e^{i(\varphi - \omega t)}$ about the equilibrium state. 
\noindent Stability of the mode depends on the sign of the imaginary part of $\omega$ (Im($\omega$)). Lopsided modes 
in the disc become unstable, in other words they are self-excited, when Im($\omega$) $> 0$ . They are stable decaying modes when Im($\omega$)$ < 0$. Modes with Im($\omega$) $= 0$ are in general stationary van Kampen modes; outside the continuum, these modes are pure normal mode oscillation of the whole disc.  


We study here the slowly varying ($\omega << \Omega$) global lopsided mode using softened self-gravity and the pressure as
 the dominant collective effect in the disc. In effect we aim to study here the lopsidedness in a real galactic disc. The effect of 
softened self-gravity was studied earlier on the slow modes in Keplerian discs by Tremaine (2001). We first neglect the collective
 effects due to the pressure in the stellar disc in order to bring out clearly the effect of self-gravity in governing the properties of
 these large-scale global modes in the galactic disc. Then we include the effect of velocity dispersion to study how the behaviour 
of lopsidedness changes in the hot disc.  
Substituting the above form of the perturbed variables into eqs.(1-3) and solving for the velocity field in the limit $\omega << \Omega$ we obtain:

$$ v_{R}^{\prime} \:=\: -\frac{i}{(1+\delta)(\omega - \omega_p)}\left[W^{\prime}_{R} + 2 W^{\prime}_{\varphi}\right] \eqno(4)$$

$$v_{\varphi}^{\prime} \:=\: \frac{1}{2 (1+\delta)(\omega - \omega_p)}\left[\delta^2 W^{\prime}_{R} + 2 W^{\prime}_{\varphi} \right] \eqno(5)$$

\noindent Where 
$$ W^{\prime}_{R} \:=\: \frac{d{\Phi^{\prime}}}{d R} + \frac{1}{\Sigma^{0}}\frac{d P_{R}^{\prime}}{d R} - \frac{P_{\varphi}^{\prime} - P_{R}^{\prime}}{R\Sigma^{0}} $$ 
\noindent
$$ W^{\prime}_{\varphi}\:=\: \frac{{\Phi^{\prime}}}{R} + \frac{P_{\varphi}^{\prime}}{R\Sigma^{0}} $$

In the above expressions $\omega_p=\Omega-\kappa$ is the free precession frequency of the $m=1$ lopsided mode and 
$\delta=\kappa/\Omega$ in the disc under consideration. Interestingly, note that the ratio of the two velocities is no 
longer a simple scalar as it is in the Keplerian limit ($\delta=1$). 

\noindent Substituting the perturbed variables in continuity equation and simplifying the equation by using the limit 
$\omega << \Omega$ we get:
  
$$ \Omega \frac{\Sigma^{\prime}}{\Sigma^{0}} \:=\: \frac{d(iv_{R}^{\prime})}{d R} - \frac{v_{\varphi}^{\prime}}{R} +  \frac{d\ln(R\Sigma^{0}) }{d R}(iv_{R}^{\prime})\eqno(6)$$

Now connecting the velocity field(eqs.[4-5]) with the continuity eq.(6) and performing straightforward mathematical 
manipulations we arrive at the following compact equation:

$$ {\omega^2}\Sigma^{\prime} + \mathfrak{D}(\Sigma^{\prime},\Phi^{\prime}) \omega + \mathcal{S}(\Sigma^{\prime},\Phi^{\prime}) \:=\: 0 \eqno(8)$$

\noindent Where
\begin{align*}
\mathfrak{D}(\Sigma^{\prime},\Phi^{\prime}) &= -\left[2\omega_p \Sigma^{\prime} + ( \mathcal{O}_{g}^{1} \Phi^{\prime} + \mathcal{O}_{p}^{1} \Sigma^{\prime}) \right]\: \: \: \: \: \: \: \: \: \: \: \: \: \: \: \: \: \: \: \: \: \: \: (9a)\\
\mathcal{S}(\Sigma^{\prime},\Phi^{\prime}) &=\omega_p^2 \Sigma^{\prime} + \omega_p ( \mathcal{O}_{g}^{1} \Phi^{\prime} + \mathcal{O}_{p}^{1} \Sigma^{\prime} )
\end{align*}
$$ \: \: \: \: \: \: \: \: \: \: \: \: \: \: \: \: \: \: \: \: \: \: \: \: \: \: \: \: \: \: \: \: \: \: \: \: \:\: \: \: \: \: \: \: \: \: \: \: \:\: \: \: \: \: \: \: \: \: \: \: \: \: \: \: - \epsilon(R) ( \mathcal{O}_{g}^{2} \Phi^{\prime} + \mathcal{O}_{p}^{2} \Sigma^{\prime} ) \: \: \: \: \: \: \eqno(9b)$$

\noindent In deriving the above eq.(8) we have used the fact that the vertically integrated pressure is a function of the surface 
density $\Sigma$. In particular we have used the perturbed pressures as $P^{\prime}_{R}=\sigma_{R}^2 \Sigma^{\prime}$ 
and $P^{\prime}_{\varphi}=\sigma_{\varphi}^2 \Sigma^{\prime}$. So we take into account of the anisotropy in pressures in the plane of the disc. In eq.(9b) $\epsilon(R)=\alpha d\omega_p/dR$ where $\alpha$ is defined below in eq.(11a). Note that $\epsilon$ would be zero if there were no differential precession in the disc.
\bigskip

\noindent We call $\mathcal{O}_{g}^{1}$ and $\mathcal{O}_{g}^{2}$ as the self-gravity operators  which are defined below:

$$ \mathcal{O}_{g}^{1}\equiv \alpha \frac{d^2}{dR^2} + \beta \frac{d}{dR} + \gamma ; \: \: \:\mathcal{O}_{g}^{2} \equiv {d}/{dR} + {2}/{R}   \eqno(10)$$

\noindent The coefficients of the operator $\mathcal{O}_{g}^{1}$ are given below:
\begin{align*}
\alpha &= \frac{\Sigma^{0}}{\Omega(1+\delta)}\: \: \: \: \: \: \: \: \: \: \: \: \: \: \: \: \: \: \: \: \: \: \: \: \: \: \: \: \: \: \: \: \: \: \: \: \: \: \: \: \: \: \: \: \: \: \: \: \:\: \: \: \: \: \: \: \: \: \: \: \:\: \: \: \: \: \: \: \: \: \: \: \: \: \: \: (11a) \\
\beta &= \frac{\Sigma^{0}}{R\Omega}\left[\frac{(\frac{d \ln\Sigma^{0}}{d\ln R} + 3 - {\delta^2/2})}{(1+\delta)} - R\frac{d\delta/dR}{(1+\delta)^2}\right]  \: \: \: \: \: \: \: \: \: \: \: \: \: \: \: \: \: (11b)\\
\gamma &= \frac{\Sigma^{0}}{R^2 \Omega}\left[\frac{(2\frac{ d\ln\Sigma^{0}}{d\ln R} - 1)}{(1+\delta)} - 2 R\frac{d\delta/dR}{(1+\delta)^2}\right] \: \: \: \: \: \: \: \: \: \: \: \: \: \: \: \: \: \: \: \: \: \: \: (11c)\\ 
\end{align*}
\noindent And we call $\mathcal{O}_{p}^{1}$  and $\mathcal{O}_{p}^{2}$ as the pressure operators which are defined below:
$$ \mathcal{O}_{p}^{1}\equiv A \frac{d^2}{dR^2} + B \frac{d}{dR} + C  \eqno(12)$$

\noindent The coefficients of the operator $\mathcal{O}_{p}^{1}$ are given below:
\begin{align*}
A &= \frac{\sigma^{2}_{R}}{\Omega(1+\delta)}\: \: \: \: \: \: \: \: \: \: \: \: \: \: \: \: \: \: \: \: \: \: \: \: \: \: \: \: \: \: \: \: \: \: \: \: \: \: \: \: \: \: \: \: \: \: \: \: \:\: \: \: \: \: \: \: \: \: \: \: \:\: \: \: \: \: \: \: \: \: \: \: \: \: \: \: (13a) \\
B &=\frac{\sigma^{2}_{R}}{R \Omega}\left[\frac{2+\delta -\delta^2/2 -\delta{\frac{d \ln\Sigma^{0}}{d\ln R}}}{1+\delta} - R\frac{d\delta/dR}{(1+\delta)^2}\right.
\end{align*}
$$ \: \: \: \: \: \: \: \: \: \: \: \: \: \: \: \: \: \: \: \: \: \: \: \: \: \: \: \: \: \: \: \: \: \: \: \: \:\: \: \: \: \: \: \: \: \: \: \: \:\: \: \: \: \: \: \: \: \: \: \: \: \: \: \: \left. + \frac{\sigma_{\varphi}^2}{\sigma_{R}^2} + \frac{2}{1+\delta}\frac{R}{\sigma_{R}^2}\frac{d\sigma_{R}^2}{d R} \right]  \eqno(13b)$$
\begin{align*}
C &= \frac{\sigma^{2}_{R}}{R^2 \Omega} ( -C_{0} + C_1 ) \: \: \: \: \: \: \: \: \: \: \: \: \: \: \: \: \: \: \: \: \: \: \: \: \: \: \: \: \: \: \: \: \: \: \: \: \:\: \: \: \: \: \: \: \: \: \: \: \:\:(13c)\\
where \\
\end{align*}

\begin{eqnarray}
\lefteqn{ C_{0}=\left[\frac{\delta +\delta^2/2 +\delta{\frac{d \ln\Sigma^{0}}{d\ln R}}}{1+\delta} + R\frac{d\delta/dR}{(1+\delta)^2}\right] }\nonumber \\
& & \: \: \: \: \: \: \: \: \: \: \: \: \: \: \: +{\frac{\sigma_{\varphi}^2}{\sigma_{R}^2}}\left[\frac{1+\delta -\delta^2/2 +\delta{\frac{d \ln\Sigma^{0}}{d\ln R}}}{1+\delta} + R\frac{d\delta/dR}{(1+\delta)^2}\right]\nonumber \: \: \: \: \: \: (14a)
\end{eqnarray}

\begin{eqnarray}
\lefteqn{ C_{1}= {\frac{R}{\sigma_{R}^2}\frac{d\sigma_{R}^2}{d R}}\left[\frac{2+\delta -\delta^2/2 -\delta{\frac{d \ln\Sigma^{0}}{d\ln R}}}{1+\delta} - R\frac{d\delta/dR}{(1+\delta)^2}\right] }\nonumber \\
& & \: \: \: \: \: \: \: \: \: \: \: \: \: \: \: \: \: \: \: \: \: \: \: \: \: \: \: \:\: \: \: \: \: \: \: \: \: \: \: \:\: +\frac{R}{\sigma_{R}^2}\frac{d\sigma_{\varphi}^2}{d R} + \frac{1}{1+\delta}\frac{R^2}{\sigma_{R}^2}\frac{d^2\sigma_{R}^2}{d R^2} \nonumber \: \: \: \: \: \: \: \: \: \: \: \: \: \: \:\:(14b)
\end{eqnarray}

\noindent The other pressure operator $\mathcal{O}_{p}^{2}$ is given by:
$$\mathcal{O}_{p}^{2} \:=\: \frac{\sigma_{R}^2}{\Sigma^{0}}\left[\frac{d}{d R} + (1+\frac{\sigma_{\phi}^2}{\sigma_{R}^2} + \frac{R}{\sigma_{R}^2}\frac{d\sigma_{\varphi}^2}{d R})\frac{1}{R}\right] \eqno(15)$$

\noindent To proceed further with the above calculations we need to know $\sigma_{R}$ and $\sigma_{\varphi}$ for the disc. 
We use the axisymmetric local stability parameter Q (Toomre, 1964) to get the radial velocity dispersion in the disc:
$$ Q \:=\: \frac{\sigma_{R}\kappa}{3.36 G \Sigma^{0}} \eqno(16) $$
\noindent And for the azimuthal velocity dispersion we use the relation $\sigma_{\varphi}/\sigma_{R}=\kappa/2\Omega$ 
(Binney \& Tremaine, 1987).
For simplicity we consider a constant Q value for the disc. 
Note that by considering a constant Q we are making the perturbed pressures arbitrary by that constant. Since we do not have an apriori knowledge about the components of the stress tensor (hence a correct pressure) from the equations of motion, we are justified in playing around with the constant Q to evalute the perturbed pressures for our analysis. Our analysis for a cold disc (Q=0) remains unaffected by such assumption. 
A proper variation of Q on the radial co-ordinate will be considered in a later work.

\bigskip

\noindent Now we require the perturbed surface density ($\Sigma^{\prime}$) and the potential ($\Phi^{\prime}$) to be connected
 through the Poisson equation in order to produce a self-consistent solution of the problem (eq.[8]). This we achieve using the
 integral form of the Poisson equation for the perturbed disc under the imposed $m=1$ lopsided mode:

$$ \Phi^{\prime}(R) \:=\: - G \int_{0}^{\infty}{d R^{\prime} R^{\prime} \mathcal{H}_{lop}(R,R^{\prime}) \Sigma^{\prime}(R^{\prime})} \eqno(17) $$

\noindent Where the kernel in eq.(17) is given by
$$\mathcal{H}_{lop}(R,R^{\prime}) \:=\: \int_{0}^{2\pi} \frac{\cos{\alpha}  d\alpha}{[R^2 + {R^{\prime}}^2 - 2 R R^{\prime}\cos{\alpha} + b^2]^{\frac{1}{2}}}  - \pi \frac{R}{{R^{\prime}}^2} \eqno(18) $$

\noindent The kernel represents the softened self-gravity of the perturbation, $b$ being the softening parameter. This allows us 
to perform the numerical integration over the nearby rings by removing the singularity at $R=R^{\prime}$. We will discuss the 
effect of this softening parameter as we progress. The second indirect term (Papaloizou, 2002) in the above kernel arises due to the $m=1$ lopsided mass 
distribution about the geometrical centre of the disc. The indirect term plays a crucial role in making the disc susceptible to the lopsided instability. The importance of the indirect term is discussed in various places in the paper. 

Substituting the perturbed potential (eq.[17]) into eq.(8) we can arive at the following equation:

$${\omega^2}\Sigma^{\prime} + \mathfrak{D}_{lop}(\Sigma^{\prime}) \omega + \mathcal{S}_{lop}(\Sigma^{\prime}) \:=\: 0 \eqno(19)$$

\noindent where
\begin{eqnarray}
\lefteqn{ \mathfrak{D}_{lop}(\Sigma^{\prime}) \:=\: -2\omega_p \Sigma^{\prime} + G\int_{0}^{\infty}{d R^{\prime} R^{\prime} \mathcal{L}_{1}(R,R^{\prime}) \Sigma^{\prime}(R^{\prime})} } \nonumber \\
& & \: \: \: \: \: \: \: \: \: \: \: \: \: \: \: \: \: \: \: \: \: \: \: \: \: \: \: \: \: \: \: \: \: \: \: \: \: \: \: \: \: \: \: \: \: \: \: \: \: \: \: \: \: \: \: \: \: \: \: \: \: \: \: \: \: \: -\mathcal{O}_{p}^{1}\Sigma^{\prime} \nonumber\: \: \: \: \: \: \: \: \: \: \: \: \: \: \: \: \: \: \:(20a)
\end{eqnarray}

\begin{eqnarray}
\lefteqn{ \mathcal{S}_{lop}(\Sigma^{\prime}) = \omega_p^2 \Sigma^{\prime} - \omega_p{G\int_{0}^{\infty}{d R^{\prime} R^{\prime} \mathcal{L}_{1}(R,R^{\prime}) \Sigma^{\prime}(R^{\prime})}} } \nonumber \\
& & +\omega_p\mathcal{O}_{p}^{1}\Sigma^{\prime} +\epsilon G\int_{0}^{\infty}{d R^{\prime} R^{\prime} \mathcal{L}_{2}(R,R^{\prime}) \Sigma^{\prime}(R^{\prime})} - \epsilon\mathcal{O}_{p}^{2}\Sigma^{\prime} \nonumber \: \: (20b)
\end{eqnarray}

The kernels in the integrals of the above eq.(20a) and eq.(20b) are given by $\mathcal{L}_{1}(R,R^{\prime})= \mathcal{O}_{g}^{1}\mathcal{H}_{lop}(R,R^{\prime})$ and $\mathcal{L}_{2}(R,R^{\prime})= \mathcal{O}_{g}^{2}\mathcal{H}_{lop}(R,R^{\prime})$. 

\noindent The above integro-differential equation (eq.[19]) describes the global behaviour of the $m=1$ lopsided mode under the 
collective effects of self-gravity and pressure in the disc. Eq.(19) can be solved by recasting it into a matrix-eigenvalue problem. By 
discretizing on a uniform grid with $N$ radial points in the disc we can write eq.(19) in a compact form:

$$ \left[I {\omega_{lop}^2} + \mathfrak{D}_{lop} \omega_{lop} + \mathcal{S}_{lop}\right] \Sigma_{lop} \:=\: 0 \eqno(21)$$

Where I, $\mathfrak{D}_{lop}$, and $\mathcal{S}_{lop}$ are the three $N \times N$ real square matrices. We call 
$\mathfrak{D}_{lop}$ as the general damping matrix and $\mathcal{S}_{lop}$ as the stiffness matrix for the galactic disc 
divided into $N$ concentric rings in analogy with the nomenclature of linear mechanical systems. Note that the damping, in 
our discrete $N$ ring system i.e. in the disc, can either be positive or negative in nature. {\it Negative damping will lead to a 
self-excitation of the discrete normal modes in the disc}. 
In the above equation $\Sigma_{lop}$ is the eigenvector corresponding to the eigenvalue $\omega_{lop}$.

\noindent The matrix elements are evaluated below: 
\begin{align*}
I_{ij}&=\delta_{ij}\\
\mathfrak{D}_{lop}^{ij} &= -2\delta_{ij} \omega_p(R_j) + \mathcal{M}_{1}(R_i,R_j)  -P_{1}(R_i,R_j) \: \: \: \: \: \:  \: \: \: \: (22a)\\
\end{align*}
\begin{eqnarray}
\lefteqn{ \mathcal{S}_{lop}^{ij} = \delta_{ij} \omega_{p}^2(R_j) - \omega_{p}(R_i)\mathcal{M}_{1}(R_i,R_j) +\omega_{p}(R_i)P_{1}(R_i,R_j) } \nonumber \\  
& & \: \: \: \: \: \: \: \: \: \: \: \: \: \: \: \: \: \: \: \: + \epsilon(R_i)\mathcal{M}_{2}(R_i,R_j) -\epsilon(R_i)P_{2}(R_i,R_j) \nonumber \: \: \: \: \: \:(22b)
\end{eqnarray}

\noindent Where $\pi \mathcal{M}_{1}(R_i,R_j)={\Delta R}R_j \mathcal{L}_{1}(R_i,R_j)$ and $2\pi \mathcal{M}_{2}(R_i,R_j) ={\Delta R}R_j \mathcal{L}_{2}(R_i,R_j)$
\noindent We use the method of finite difference to evaluate the contribution to the matrix elements from the pressure terms e.g.$P_{1}(R_i,R_j)$ and $P_{2}(R_i,R_j)$. 


\subsection {Technique to solve the equations}
Eq.(21) represents an $N$-dimensional quadratic eigenvalue problem(QEP) in the eigenvalue $\omega_{lop}$ and $N$-dimensional
 eigenvector $\Sigma_{lop}$ describing the lopsided perturbation surface density in the disc. We consider $N=100$ rings uniformly 
spaced up to an outer boundary $R_{out}$ which we keep as a free parameter in our study. Since the original eq.(19) is an 
integro-differential equation we need to supply boundary conditions to solve the matrix problem. We use the Neumann boundary
 conditions $d\Sigma^{\prime}/dR=0$ on the boundary in order to evaluate the contributions to the matrix elements from the 
pressure terms. Note that in the absence of velocity dispersion i.e. for a cold disc we have a simple integral equations. The above boundary condition is used only to solve the govering equation in a hot disc. For a cold disc, we do not need to impose any external boundary condition; the boundary condition is in-built in the integral equation. The standard 
way to solve the QEP (eq. [21]) is to reduce it to a generalized eigenvalue problem(GEP). We use LAPACK subroutines for the 
numerical solution of our GEP. For more numerical details see Saha \& Jog (2006) and relevant references therein.
 
\section {Results}
We study the lopsided mode in a galactic exponential disc with the central surface density $\Sigma_{0}$ and a scale-length $R_d$:
$$\Sigma^{0}(R) \:=\: \Sigma_{0}e^{-R/R_d} \eqno(23)$$
\noindent as observed in most spiral galaxies (Freeman 1970).
We limit our investigations up to $4$ disc scale-lengths, typically within the optical region, which contains $\sim 90\%$ of the total stellar disc mass. 
We do not include the effect of the dark matter halo in the present problem 
since different dynamical mechanisms will then be involved, and we want to
examine them in turn.

\noindent In all the calculations we get the perturbation surface density i.e. eigenvector $\Sigma_{lop}$ (eq.[21]) in a 
dimensionless form and the eigenvalues $\omega_{lop}$ corresponding to the eigenvector are in units of $\sqrt{{\pi G \Sigma_{0}}/R_d}$. For disc like our Milky Way the value of this unit $\sim 52.9$ km$s^{-1}$kpc$^{-1}$. 

\noindent We first consider here a cold self-gravitating disc i.e a disc with zero velocity dispersion. The unperturbed disc is 
rotationally supported against the gravitational attraction. The underlying motivation is to find out how the behaviour of 
lopsidedness ($m=1$ mode) in a differentially rotating cold self-gravitating disc changes when a finite non-zero velocity 
dispersion is included in the disc. Below we report the results first for a cold disc and subsequently the results from the hot disc are discussed.
 
\subsection {\bf{Lopsidedness in a Cold Disc :}}
We show that the cold self-gravitating discs are able to support $m=1$ discrete normal lopsided modes. These modes are important 
because they show the observational signatures and they also turn out to be long-lived as discussed below.

\subsubsection {\bf {Lopsided mass distribution in stellar disc}}
\begin{figure}
{\rotatebox{270}{\resizebox{5cm}{8cm}{\includegraphics{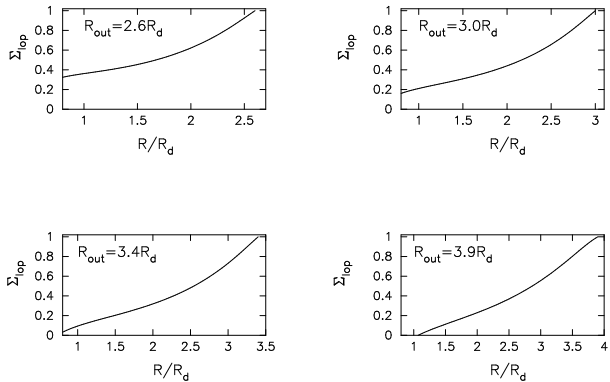}}}}
\begin{center}

\caption{Behaviour of lopsided modes in a cold disc. Mode shapes are quite robust with respect to the changes in the disc size. 
The amplitudes of these modes are in arbitrary units.}

\end{center}
\end{figure}

In Fig. 1 we have shown four subplots of the lopsided modes at various values of the outer boundary $R_{out}$. The 
$N$-dimensional matrix eigenvalue equation (eq.[21]) produces $2N$ eigenvalues and $2N$ eigenvectors in the disc. 
Observationally relevant modes are actually the lowest frequency ground state modes. By ground state we mean the mode with the lowest value of the real part of the eigenvalue (Re($\omega_{lop}$)) in the eigen spectrum. These modes are with the lowest number of nodes, typically zero or one node at most, showing the global nature of the lopsidedness in the disc. For smaller sized disc (as shown in the first two panels in fig. (1)) these modes are very interesting in nature, they are discrete and distinct (as can be seen in fig. (4)), in the sense that these modes represent a very slowly precessing stationary lopsided pattern. The imaginary parts of the eigenvalues corresponding to these pattern are zero i.e. Im($\omega_{lop}$)=0 for these modes. However as we increase the disc size this scenario changes and collective effect due to disc self-gravity becomes very important. We find that for a comparatively larger disc the most slowly precessing pattern appears as an instability in the system. The reason for this instability lies in the collective effect due to the softened self-gravity. Even though its hard to exactly pinpoint the source of the lopsided instability, our numerical analysis show that without the indirect term the ground-state lopsided modes are never unstable in the disc.  

 We have used the value of softening parameter $b$ $\sim$ some fraction of the inter ring spacings in order to solve the matrix eigenvalue problem. We show that the mode shape corresponding to the lowest pattern speed is quite robust with respect to the variations in both the parameters namely $R_{out}$ and $b$. 
In Figs.1a - 1d,  we have varied $R_{out}$ from $2.8 R_d$ to $3.8 R_d$ and clearly in all these cases the mode amplitudes 
increase with the galactocentric radius, in a good agreement with the observations (Section 1).
 
The surface density contours for the lopsided disc : $\Sigma(R,\varphi,t)=\Sigma^{0}(R) + \delta_p \Sigma^{\prime}(R,\varphi,t)$ are 
shown in Fig. 2. We have considered $\delta_p=0.02$ and the contours are at t=0. This clearly brings out that the contours deviate 
from the unperturbed disc surface density more at larger radii. 
A barycentric shift in the mass distribution of the disc due to the lopsided mode is obvious from fig.2. Such a displacement of the center of mass of the disc due to a one-armed spiral mode is discussed in great detail in Evans \& Reed, 1998. They show that a one-armed spiral does not shift the barycentre of the disc substantially. The lopsided mode in our analysis seems to precess with a very slow pattern speed (as can be seen in fig.3 and fig.8). Because of this slow pattern of oscillation we expect the barycentre of the disc to be shifted only by a small amount and the mass distribution of the disc does not alter appreciably; it probably stays in that state for a long time. However, since in the linear regime the amplitude of the lopsided surface density is arbitrary, we do not think it is possible to calculate the exact shift of the barycentre due to the lopsided pattern.      
\begin{figure}
{\rotatebox{270}{\resizebox{5cm}{6cm}{\includegraphics{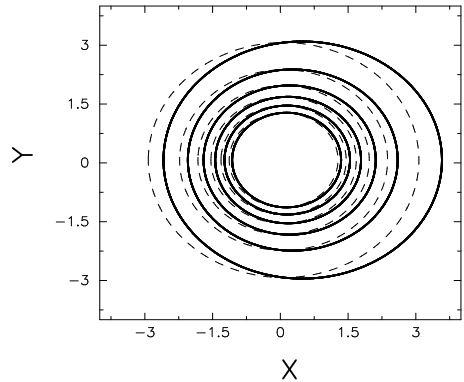}}}}
\begin{center}

\caption{Surface density contours of the lopsided disc for $\delta_p=0.02$. The maximum of the surface density occurs at (0,0). 
The x and y axes are in units of $R_d$. The outward contours are more and more deviated from the unperturbed circular
 contours --- showing that lopsided behaviour is prominent in the outward direction. The contour levels are 0.05, 0.1, ...0.3 times 
the central surface density.}
\end{center}
\end{figure}

So our study - based on the internal disc dynamics- naturally explains the observational fact that the lopsidedness increases 
outward in the disc.

\begin{figure}
{\rotatebox{270}{\resizebox{5cm}{7cm}{\includegraphics{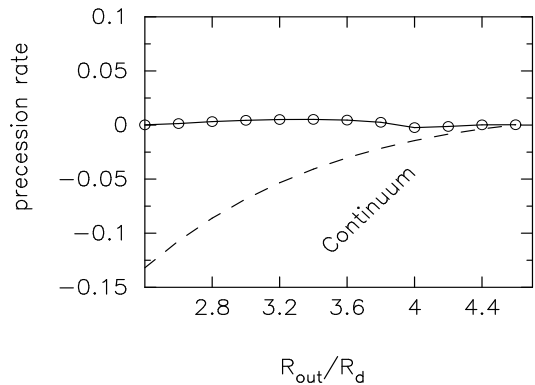}}}}
\begin{center}

\caption{The solid line, connecting the open circles, shows the precession rate $Re(\omega_{lop})$ of the lopsided mode under the collective effect 
of the self-gravity. The dashed line is free precession and acting as the lower boundary ($\Omega -\kappa$) of the continuum. 
The upper boundary ($\Omega +\kappa$) is not shown because it is orders of magnitude higher than $Re(\omega_{lop})$. }

\end{center}
\end{figure}

\subsubsection {\bf{Persistence of the lopsided mode}}
In order to achieve a persistent lopsided pattern in the disc, the pattern formed initially  by the stellar orbits should be synchronized 
in such a way that they rotate with a constant pattern speed. In the kinematical picture this requires that the differential precession in 
$\Omega - \kappa$ must be zero which is true in a Keplerian disc 
because the disc self-gravity is insignificant and the lopsided pattern would remain intact there for all time. 
In a sharp contrast, in a galactic stellar disc, it is the self-gravity which is the most important. So the survival of lopsidedness in the 
stellar disc is puzzling. The strong differential precession in $\kappa -\Omega$ in the inner region would wash away any coherent pattern imposed onto it.
The typical winding up time scale for the lopsided pattern can be obtained by using the formula $\tau_{lop}=2\pi/{\Delta(\kappa - \Omega)}$ (See BLS). Their estimation shows that the typical lifetime of the lopsided pattern seen in outer HI disc is $\sim 1-5 $ Gyr. 
For a typical stellar disc like that of our Galaxy this winding up time scale in the optical region would be less than a Gyr. So in 
the stellar disc the survival problem of the lopsidedness is far more severe. 

In Fig. 3 we have shown the variation of the $Re(\omega_{lop})$, denoting the pattern speed of the lopsided mode in the 
exponential disc, with respect to the size of the disc. The eigenfrequencies $\omega_{lop}$ are corresponding to the
 ground-state lopsided mode in the disc. We see that $Re(\omega_{lop})$ is almost constant compared to the variation in 
$\Omega - \kappa$ and thereby the disc avoids the differential precession. In this way the winding-up time scale increases 
by a factor of $\sim$ 10 or more and this turns out to be more than 12 Gyr for a disc like that of our Galaxy. This implies that the lowest frequency modes are almost non-rotating pattern of the disc and N-body simulations (as described later in sec.4) confirms this picture. Note that in all our calculations 
$Re(\omega_{lop})$ lie in between the gap made by $\Omega - \kappa$ and $\Omega + \kappa$ which act as the lower and 
upper boundary of the continuum region respectively. In other words we have $\Omega - \kappa <Re(\omega_{lop})<\Omega + \kappa$. 
All the ground-state lopsided modes in our calculations are thus discrete in nature; once $Re(\omega_{lop})$ lies outside this gap 
they fall in the continuum regime. We show that as the disc size increases the width of the gap shrinks and eventually any discrete 
lopsided mode ceases to exist! And this phenomenon occurs at a region near 4.6 disc scale-lengths where $Re(\omega_{lop}) \sim {\kappa-\Omega}$ (see fig.3). Surprisingly there is a natural resonance at $\kappa=\Omega$ at around 
4.6 disc scale-lengths. So we see that vanishing of the discrete lopsided modes occurs  below this resonance and 
this might be acting naturally as a boundary beyond which no discrete lopsided mode can exist in a purely exponential disc.  
                     
\noindent Fig. 3 also brings out another fact clearly regarding the sense of precession of the lopsided pattern in the cold disc. In linear analysis the final equation (21) is invarinat under the transformation $\omega_{lop}\rightarrow -{\bar{\omega}_{lop}}$, one would expect the pattern speed to be both retrograde and prograde. Our numerical eigenvalue analysis shows that both the retrograde and prograde pattern are likely to occur in the cold disc. some of the earlier works namely Statler (1999) has argued that the precession of a self-consistent lopsided disc must be prograde; 
Jacobs \& Sellwood (2001) find in their simulation that the lopsided modes precess in prograde direction. 


\subsubsection{\bf {Complete eigen spectrum of the lopsided pattern in a cold disc}}
On solving the N-dimensional eigenvalue problem (eq.[21]) in the case of cold discs (Q=0) with different disc sizes, we obtain 2N complex eigenvalues and these complex eigenvalues are plotted in the {\it{Argand}} diagram as shown in fig.4. The complete eigen spectrum in the case of spiral and bar-like structures in galaxies are discussed by Polyachenko, 2004. The complex eigenvalues appear as complex conjugate pairs in the full spectrum. In the Argand diagram, we will concentrate only to those points for which Re($\omega_{lop}$) $\sim$ 0 since we are primarily interested in the slow modes. For smaller sized disc as in panel-a, there exists a distinct (clearly separated from the rest of the points in the complex plane) point in the discrete region close to the boundary of the lower continuum (denoted by the vertical solid line) corresponding to Re($\omega_{lop}$) $\sim$ 0 and Im($\omega_{lop}$)=0. The mode corresponding to this point is an almost non-rotating stationary global lopsided pattern of the disc. Being isolated in the Argand diagram {\it{this mode resembles like an excited normal mode of the disc}} much like what Lynden-Bell in 1965 first conjectured in the context of warps (basically $m=1$ mode in the perpendicular direction to the disc midplane) that they are excited normal modes of the discs. The existence of such a mode of oscillation outside the continuum was shown by Mathur, 1990 for one dimensional and 3D gravitating system. {\it{The presence of such a distinct normal mode in our eigen solution for a disc-like 2D self-gravitating system confirms the previous trend found in theoretical studies namely self-gravitating system supports normal modes of oscillation in the disc}}.  

Interestingly, note the appearance of a complex conjugate pair of eigenvalues in the continuum region near the solid line in panel-a. As the disc size increases this pair drifts towards the discrete region from the continuum and at around 4 disc scale lengths, they are in the discrete region. And the slowly  precessing stationary pattern turns out to be a lopsided instablity in the cold disc. This instability, in comparatively larger discs, arises due to the collective effect of the softened self-gravity of the perturbation where the indirect term place a crucial role in driving the instability. Whereas for smaller sized discs, which are like a broad annular ring, the collective effect of softened self-gravity is not enough to drive an instability and the lopsided pattern basically represents a stationary oscillation of the whole disc.   

\noindent Apart from the discrete region, the complete eigen spectrum contains a lot of modes in the continuum region. For a cold disc only a few of these modes are with Im($\omega_{lop}$)=0 representing stationary oscillation in the system. These particular normal modes (as shown in fig.5) are interesting because they resemble the van Kampen modes in plasma physics. These modes would probably decay by Landau damping and hence are of no interest in our present study.   

\begin{figure}
{\rotatebox{270}{\resizebox{5cm}{8cm}{\includegraphics{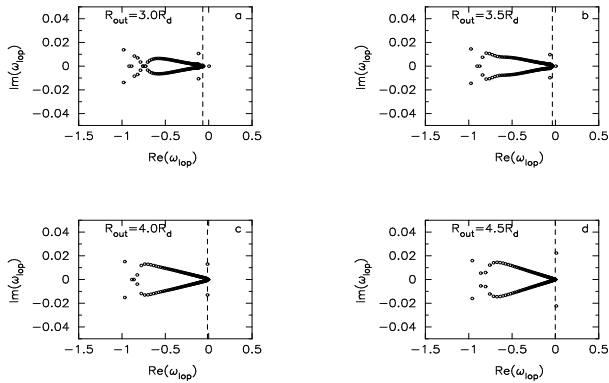}}}}
\begin{center}

\caption{A plot showing the behaviour of the complex eigenvalues corresponding to the lopsided modes in a cold disc (Q=0). Each panel shows the complete eigen solution of eq.[21] for a different value of the disc size $R_{out}$. The vertical dashed line is the lower boundary of the continuum as shown in fig. 3 for each disc size. }

\end{center}
\end{figure}


\begin{figure}
{\rotatebox{270}{\resizebox{5cm}{8cm}{\includegraphics{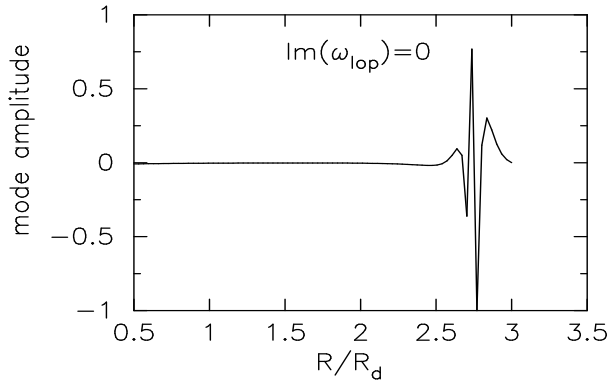}}}}
\begin{center}

\caption{The plot showing the typical behaviour of a stationary van Kampen mode lying in the continuum for Rout=3.0. These continuum modes are corresponding to Im($\omega_{lop}$)=0.}

\end{center}
\end{figure}


\subsection{\bf {Lopsidedness in a hot disc :}}
In reality galactic discs are never cold. Because of a non-zero velocity dispersion a hot disc is supported by rotation as 
well as pressure against the gravitation. We quantify the hotness of the discs by the Toomre's Q (eq.[16]) parameter; 
note that Q=0 refers to a cold disc and Q$\neq$0 means a hot disc. If Q is less than the critical value of 1, the disc is 
axisymmetrically unstable. Basically Q serves as a very useful local criterion for the axisymmetric stability of the disc. 
Normally Q has a radial profile in the disc. For the sake of simplicity we consider Q=constant throughout the disc. 
By doing this we would like to understand what happens to the lopsided perturbations when there is a non-zero Q in the disc.  
 
\subsubsection{\bf {Complete eigen spectrum of the lopsided pattern in a hot disc}}
The behaviour of the cold disc against lopsided perturbation changes dramatically as soon as a finite velocity dispersion is introduced in the system. The transformation in the disc behaviour is best seen through the complete eigen spectrum plotted (fig.6) in the Argand diagram for different values of Q. In the case of Q=0, there are only a few modes in the continuum with Im($\omega_{lop}$)=0 and only one or two in the discrete region. Most of the modes in the continuum are of unstable in nature. However as we increase the Q value, the disc changes abruptly and modes in the continuum are populated along the real line i.e. with Im($\omega_{lop}$)=0. For Q=0.2 in this case shows that there are almost no unstable modes in the discrete region, some of them are indeed in the continuum with high pattern speeds. At Q=0.6, we see the appearence of unstable modes in the discrete region of the spectrum. 
The appearence of the unstable mode in the spectrum is not very obvious. There are subtle interplay amongst the indirect term in the poisson equation (eq.[18]), the softened part of the self-gravity and the velocity dispersion (Q). Without the indirect term the ground state modes are never unstable, no matter what the values of the softening ($b$) and the Q-value in the disc. The indirect term turns out to be the main driver of the lopsided instability in the disc. 

Beyond this as Q increases, the number of unstable modes goes down and at some Q=Q$_{crit}$ no unstable modes are found with lowest pattern speed. Beyond Q$_{crit}$ the modes with lowest value of the pattern speed in hot disc again turn out to be purely oscillatory.

\subsubsection{\bf {Dependence of the growth rate on velocity dispersion}}
We solve the governing eq.(21) for the lopsided modes in a hot disc with non-zero radial ($\sigma_{R}$) and azimuthal 
($\sigma_{\varphi}$) velocity dispersions. Inclusion of velocity dispersion has an effect opposite to that of the self-gravity 
in the disc. In a cold disc, the imposed lopsided perturbation emerges to be a stationary normal mode oscillation for smaller sized disc or grows as an instability when the collective effect of the softened self-gravity is sufficient as it is in comparatively larger sized discs. By addition of a finite non-zero velocity dispersion (quantified here by a constant value 
of the Q-parameter) in the disc the strong effect of self-gravity is diluted. As we increase the Q-value the growth rate 
($\gamma_{lop}$=Im($\omega_{lop}$)) decreases and there exists a Q$_{crit}$ beyond which we find that the lowest frequency modes are no 
longer unstable in the disc, see Fig.7. Note that as we increase the Q-value, the growth rate reaches a maximum and starts 
to decline; a similar behaviour of the growth rate vs Q was observed in the global analysis of $m=1$ mode in a nearly Keplerian 
disc (Adams et al. 1989). In Fig.7 we find the value of Q$_{crit}$ to be 1.0 for a disc with R$_{out}$=3.5 R$_d$, for other 
values of R$_{out}$ we find that Q$_{crit}$ normally lies around the 1. The value of Q$_{crit}$ may be surprising to the 
reader; note that we are considering here a purely stellar disc without any dark matter halo. Take for example the stellar disc of our 
Galaxy with central surface density $\Sigma_{0}$ = 640 M$_{\odot}$pc$^{-2}$, R$_d$=3.2 kpc and the radial velocity dispersion
 ($\sigma_{R}$) as according to the Lewis-Freeman law (Lewis \& Freeman 1989), the Toomre's Q parameter is always less than 
1 beyond 1 disc scale length.  

\begin{figure}
\rotatebox{-90}{\includegraphics[height=8cm]{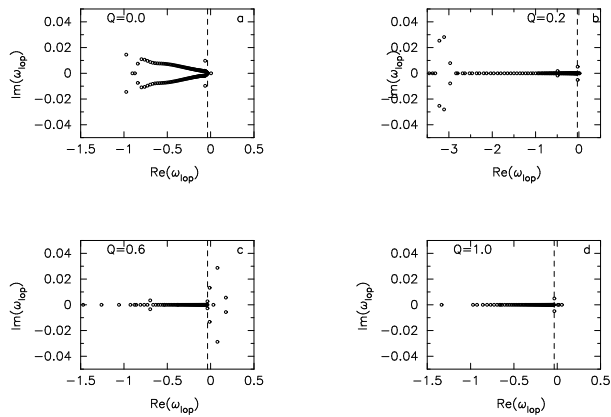}}
\caption{The full eigen spectrum for a disc at different values of Q, including one with Q=0.0. The figure shows how the eigen spectrum behaves in the Argand diagram as finite velocity dispersion is introduced in the cold disc of size $R_{out}=3.5 R_d$. The verticval dashed line represents the lower boundary of the continuum as shown in fig. 3 for the disc.}
\end{figure}

\begin{figure}
\rotatebox{-90}{\includegraphics[height=8cm]{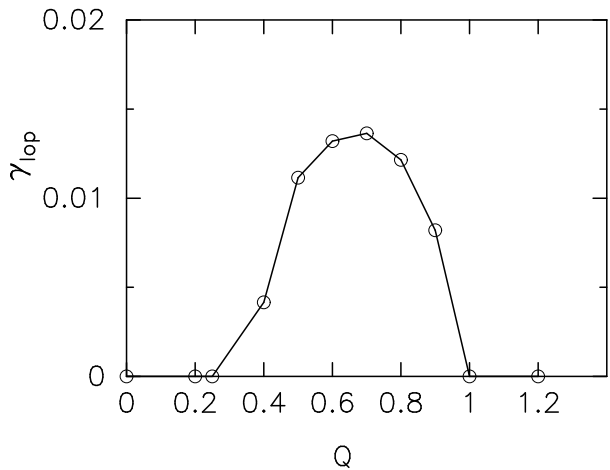}}
\caption{Growth rates ($\gamma_{lop}$) of the lowest frequency modes are plotted against the Q-parameter. The growth rate 
reduces as Q increases in the disc followed by a peak. Beyond Q$_{crit}$ these lopsided modes are not unstable anymore.}
\end{figure}

\subsubsection{\bf {Dependence of the lopsided pattern on velocity dispersion}}

In Fig.9 we study the behavior of the lopsided modes in a hot disc. These lopsided modes are again the lowest frequency (i.e. the lowest 
value of the real part of the complex eigenvalue) eigenmodes out of the 2N eigenmodes in the N-ring eigen-system. In a cold disc these lowest frequency eigenmodes are interesting or have observational relevance because they are global in extent (having zero or one radial node). With the introduction of a finite velocity dispersion, the behaviour of the eigenmodes also changes drastically. In Fig.9 the solid line is the eigen mode in a cold disc i.e. Q=0. As Q-value increases in the disc, the self-gravity starts falling down and fails to 
synchronize the adjacent rings to precess them coherently and produce a global mode. The eigen modes start acquiring radial nodes. The number of radial nodes increases as Q approaches the value Q$_{crit}$. At around Q$_{crit}$ 
the eigen modes have vanishing amplitudes with many radial nodes beyond 1 disc scale length showing that there is no coherent lopsided pattern anymore in the disc beyond Q=Q$_{crit}$. 
In fig.8 we show the variation of the pattern speed with the Q-value. Interestingly the pattern speed does not get affected in appreciable amount with the increase in Q. The pattern speeds of the ground state modes are  not quite insensitive to the increment in the velocity dispersion. Our numerical calculation show that there is no definite sense of precession of the lopsided pattern in the disc. Fig. 8 also brings out clearly another fact that the global lopsided pattern precesses quite slowly even in a hot disc! Even the dynamical friction (Ideta, 2002) from an external dark matter halo would take long time to damp out such a pattern in the galactic disc.    
        
\begin{figure}
{\rotatebox{270}{\resizebox{5cm}{8cm}{\includegraphics{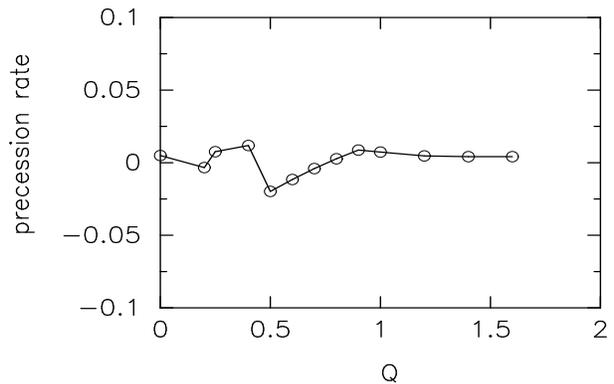}}}}
\begin{center}

\caption{A plot of the pattern speed vs the Q value in the disc for a disc size of 3.5 $R_d$. Clearly, there is no preferred sense of precession of the lopsided pattern in the disc.}  

\end{center}
\end{figure}

\begin{figure}
\rotatebox{-90}{\includegraphics[height=8cm]{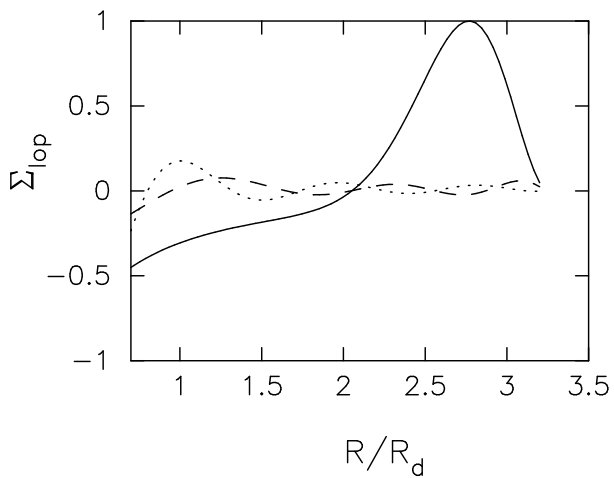}}
\caption{The lowest frequency modes are plotted against the radius for different values of the Q-parameters at t=0. As Q increases in the disc, the number of radial nodes of these modes increases. The solid curve is for Q=0.2; dashed one is for Q=0.6 and the dotted one is for Q$_{crit}$=1.0.}. Note that when Q=0.0 i.e. for a cold disc, the lopside modes are coherent with zero or one node at most (
fig. 1). This coherence in the lopsidedness is lost as the disc becomes hotter.
\end{figure}

\section{N-body simulations}
\label{nbody}

  In order to test the predictions of the previous analysis, numerical
simulations have been performed. Since different physical mechanisms
can occur in the presence of dark matter halo, and the presence of 
a gaseous component, the present work is restricted to a pure stellar disc.
Initially there is no bulge component, although a pseudo-bulge is building
along  the simulations, through bar instability, which then triggers a
box-peanut shaped bulge formation.

These simulations are meant to isolate the dynamical mechanisms
occurring in a pure stellar disc, for academic purposes. In a following
paper, more realistic galaxies with dark matter haloes, and gaseous
components will be considered.

\subsection{Numerical details}

The galaxy is isolated, and represented by only stellar
particles. The most efficient N-body algorithm to simulate such a
distribution is a PM (particle-mesh) code.  Since in particular
$m=1$ modes and lopsidedness are of interest, polar grids
with higher spatial resolution in the center are not adapted: the high 
density core of the galaxy is expected to move around, and 
the variable resolution could produce artefacts. A cartesian grid
is therefore selected. We use here an
 FFT N-body code with the James (1977) method to avoid the influence
of Fourier images. The useful grid is 256$^2$x128, corresponding to 
(60 kpc)$^2$x30 kpc, implying  a cell size of 234 pc.
This size is also the softening length of the Newtonian gravity. The total number
of particles is between 8 10$^5$ and 1.6 10$^6$, according to the various runs. 

The initial radial distribution of particles is not 
very important, since it is known that a pure disc without
any spherical component (such as bulge or halo) is unstable 
with respect to bar formation; then the bar re-distributes
efficiently the particles, to form an exponential disc (e.g. 
Combes et al 1990, Pfenniger \& Friedli 1991). 
For the sake of simplicity, 
the initial distribution of stellar particles is taken 
according to a Kuzmin-Toomre disc, of surface density:
$$
\Sigma(r) = \Sigma_0 (1 +r^2/r_d^2)^{-3/2}
$$
where $r_d$ is the characteristic scale-size of the disc,
truncated at 25kpc, with a disc mass M$_d$.
The initial velocity distribution of the stars in the disc
is computed from the stationary Jeans equation in cylindrical 
coordinates, yielding the asymmetric drift $v_{rot} - v_{cir}$,
assuming 
that $ \sigma_\varphi/\sigma_r = \kappa/(2 \Omega) $,
as in the epicyclic approximation.

  The distribution of density perpendicular to the disc is that of
an isothermal thin layer:
$ \rho(z) = \rho_0  sech^2 (z/H) $
with a height slightly flaring with radius, as
$H(r) = H_0 (1 +(r/9kpc)^2)^{0.35} $,
where $H_0 = 700 pc$. The z-velocity dispersion is then
computed from 
$\sigma_z^2 = H(r)\, \pi \, G \, \Sigma(r)$.
 
The initial Toomre Q parameter is slightly varying with radius as:
$ Q = \sigma_r/\sigma_{crit} = Q_0/exp[(r_d/23kpc)^2]$
The time step is 1 Myr. The parameters of the runs described here are
displayed in Table \ref{param}.

\begin{table}[h]
\caption[ ]{Parameters of the initial conditions}
\begin{flushleft}
\begin{tabular}{cccccc}  \hline
Run    & r$_d$   &  M$_{d}^*$  &   N$_{par}$    &  Q$_{0}$ \\
       & kpc     &  M$_\odot$  &                &          \\
\hline
Run A  &  3.5  &  4. 10$^{10}$ &  0.8 10$^{6}$ &   2.3     \\
Run B  &  3.5  &  4. 10$^{10}$ &  1.6 10$^{6}$ &   2.3    \\
Run C  &  3.5  &  4. 10$^{10}$ &  1.6 10$^{6}$ &   1.2    \\
Run D  &  1.8  &  4. 10$^{10}$ &  0.8 10$^{6}$ &   2.3    \\
\hline
\end{tabular}
\end{flushleft}
$^*$ mass inside 25 kpc radius\\
\label{param}
\end{table}

\begin{figure}
\rotatebox{-90}{\includegraphics[height=8.5cm]{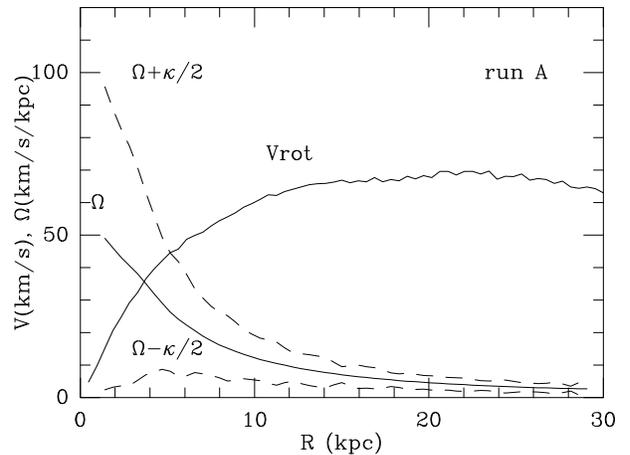}}
\caption{Rotation velocity $V_{rot}$ of the stars, at the
end of simulations (run A, T=12.6 Gyr), and frequencies $\Omega$,
$\Omega-\kappa/2$ and $\Omega+\kappa/2$ derived from
the potential at this epoch for run A.}
\label{vrot}
\end{figure}

The stellar rotation curve corresponding to  the reference run A
is plotted at the end of the simulation in Fig.~\ref{vrot}.
The characteristic frequencies $\Omega$, $\Omega -\kappa/2$
and $\Omega +\kappa/2$ are derived from the measured potential,
at the final epoch.

\subsection{Results}

The behaviour of run A and run B is quite similar. Run B has
twice more particles, and demonstrates how the particle noise 
may affect the estimation of the $m=1$ amplitudes (error bars will be given below).

\begin{figure}
\rotatebox{-90}{\includegraphics[height=8.5cm]{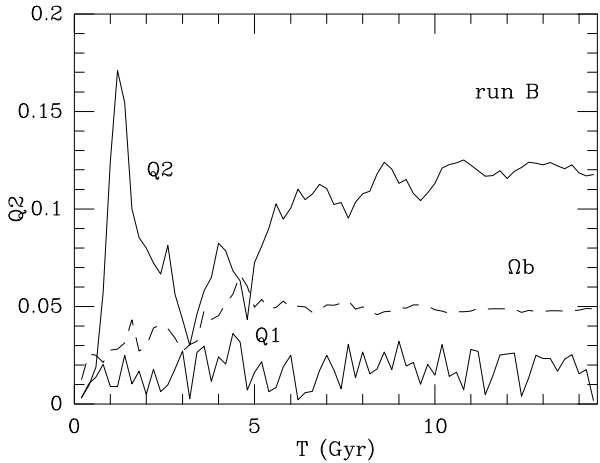}}
\caption{ Strength of the $m=2$ mode ($Q_2$) and pattern speed $\Omega_b$, for
run B. Units of $\Omega_b$ are 100km/s/kpc. Also shown is the 
$m=1$ strength $Q_1$.}
\label{bar-runB}
\end{figure}

Our reference run (A or B) reveals a bar instability as expected,
given that no dark matter halo, nor bulge is present to stabilise
the exponential disc.
The strength of the bar is estimated by the ratio of the tangential
force of the $m=2$ Fourier component, normalized to the radial force.
If the potential  is decomposed as
\begin{equation}
\Phi(r,\varphi) = \Phi_0(r) + \sum_m \Phi_m(r) \cos (m \varphi - \phi_m)
\end{equation}
the bar strength is the maximum over radius of
\begin{equation}
Q_2 = 2 \Phi_2 / r | F_r |
\end{equation}
where $ F_r = -\partial \Phi_0 /  \partial$r.
  The evolution of $Q_2$  for run B is shown in
Fig \ref{bar-runB}, as well as its pattern speed
$\Omega_b$, measured from the monitoring of the phase of the $m=2$
pattern with time.

In the beginning of the evolution, the  $m=2$ strength has a contribution
from the spiral structure which develops, and helps to transfer angular
momentum in the outer parts. After a few Gyrs, and in the quasi-stationary
state, there is only a bar. 

\begin{figure*}
\includegraphics[height=17cm,angle=-90]{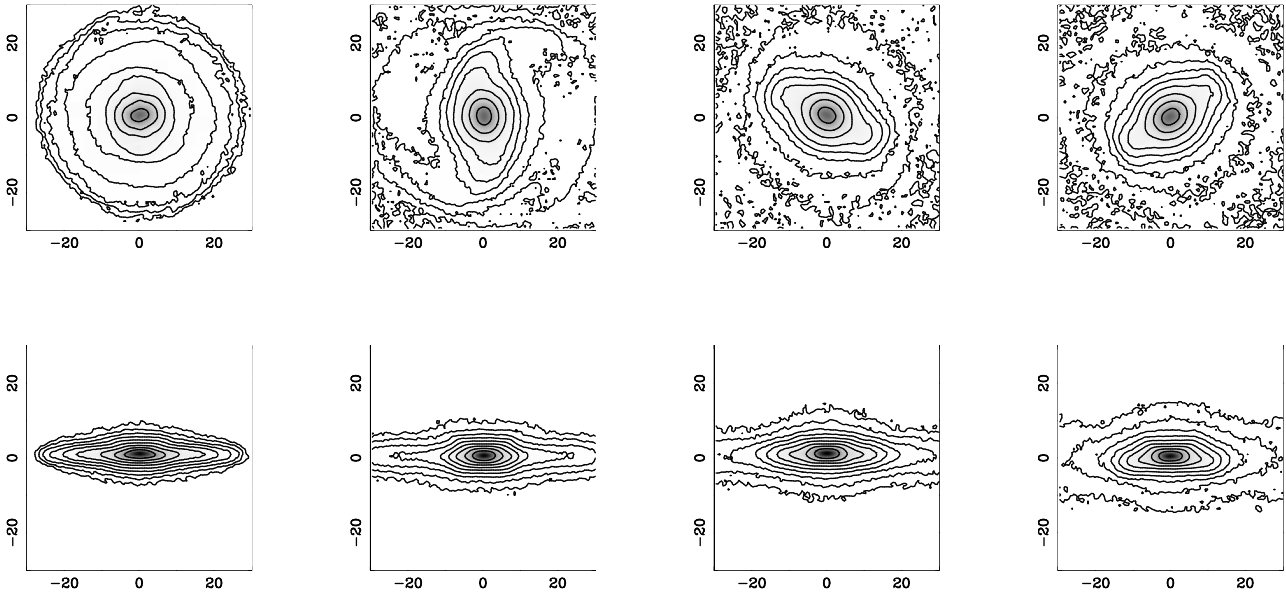}
\caption{ Run B: Contours in logarithmic scale of the surface density of
the stellar disc, face-on (top) and edge-on (bottom), at four different
epochs: T=0.8, 5.6, 9.6, 14.4 Gyr, from left to right.  The contours are taken between
the maximum surface density, and a minimum equal to 5 10$^{-4}$ that value. In units of
900 M$_\odot$/pc$^2$, in square pixels of 470pc in size, the maxima are respectively
0.6,0.6,0.6,0.6 for the face-on values, and 1.6, 1.9, 1.5 and 1.5 for the edge-on values.}
\label{cont-runB}
\end{figure*}

The bar instability heats the stellar disc, which 
in a first phase weakens the $m=2$ pattern that developped its
maximum around 1Gyr. Then the bar continues to grow slowly
until its strength reaches a stationary value, of $Q_2 \sim 0.1$.
  The bar pattern speed increases slowly in this evolution,
which is due to the strong concentration of matter driven
by the bar. The face-on morphology of the galaxy is displayed 
in the contours of Fig \ref{cont-runB}. The edge-on projections
reveal the build-up of a peanut morphology, which considerably
thickens the stellar disc, and forms a pseudo-bulge.

The abrupt fall-off of the  $Q_2 $ strength in the first evolution
phases (after 2-3 Gyr), is due to the violent instability of the beginning,
occuring so shortly with respect to the dynamical time, that 
the family of orbits sustaining the bar have no time to re-arrange,
before the radial redistribution of matter. The mass is re-distributed radially,
in consequence of the angular momentum flow towards the outskirts.
The exponential disk length decreases, and the concentration of matter 
has the consequence that the bar pattern speed  increases.
Then the resonances change radial location, and therefore involve different
particles, which means that the bar can now grow again with 
a different pattern speed, and different sustaining orbits.

\begin{figure}
\rotatebox{-90}{\includegraphics[height=8.5cm]{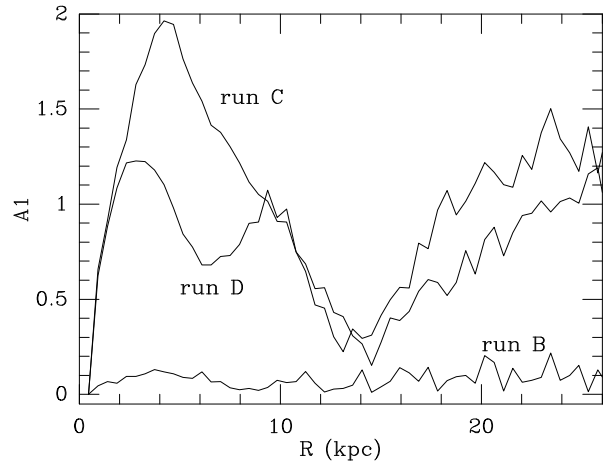}}
\caption{Radial behaviour of $A_1$ for all runs, at the epoch
T= 7.2 Gyr. $A_1$ is the $m=1$ Fourier component of the
surface density, normalised with the $m=0$ component.}
\label{a1-toutr}
\end{figure}

To study the simultaneous development of any $m=1$ mode,
the analogous $Q_1$ term was computed from the Fourier analysis
of the potential. However, the $m=1$ amplitude is weaker, and 
the estimator from the potential, which is a more global way
to quantify the lopsidedness, dilutes any local maximum, and 
therefore revealed too noisy. An estimation from the density map
in the face-on projection of the galaxy, was then preferred.
This is an estimator currently chosen by observers
(e.g. Zaritsky \& Rix 1997, Bournaud et al 2005).
 From the Fourier analysis of the surface density in
the plane, we define $A_1$ the ratio of the $m=1$
and $m=0$ term. The behaviour of $A_1$ as a function 
of radius is displayed for all runs in Fig \ref{a1-toutr}.
For run C and D, there is a strong maximum in the center
 of the disk (around 5 kpc). After a pronounced minimum around 14kpc,
 the amplitude of A1 is rising again in the outer parts.
 The projected density maps of the perturbed disks show
 that this $m=1$ is essentially due to the off-centering of the corresponding
 parts of the disk. The central parts are shifted on one side, while
 the outers parts are shifted on the opposite side. This explains
 the minimum in the intermediate radii (around 14kpc).
 Although there is mainly at the beginning an $m=1$ spiral
 structure, the dominating feature of the $m=1$ mode is the
 off-centering.

It can be noted that the estimator gives a coherent
value getting out of the noise between 2 and 10kpc.  Values at large radii are always
noisier, due to the low number of particles, in the outer parts of an exponentially decreasing surface density. 
In the very center, the small area involved also does not include enough particles.  We therefore select 
for all runs the common radius R= 5kpc to estimate $A_1$ and deduce the 
time variations.

While Run B is  unstable to the $m=2$ mode (bar and spiral),
it does not show a strong $m=1$ perturbation, which could be expected,
given the high initial velocity dispersion ($Q=2.3$). 
Run C on the contrary
is violently unstable with respect to $m=1$, as can be seen in the 
evolution of pattern strength in Fig \ref{bar-runC} and in the
contour plots of Figure \ref{cont-runC}.
This is easily explained through the lower initial velocity dispersion $Q=1.2$.

Both runs develop a bar however, and are unstable
with respect to the buckling instability, forming the now
well-known peanut shape (e.g. Combes et al 1990, 
Martinez-Valpuesta et al 2006). The peanut formation
interferes with the $m=1$ mode to make the z-elevation
of particles asymmetric, and produce a marked tilt in
the disc plane at the end of the simulations. 
This does not occur for run B, where the distribution of
particles remain roughly symmetrical with respect to the 
center of mass. But Run C shows very well this phenomenon.
In the buckling some particles are elevated upwards, at some
radii, in resonance with the vertical Lindblad resonance, which
happens to coincide with the in-plane ILR, since $\kappa \sim \nu_z$,
at these radii. The lopsidedness happens to give more weight to the population
of particles that  are elevated upward, breaking the symmetry with respect
to the plane. Some particles escape at larger radii, and are lost for
the equilibrium. The remaining disc re-settle in a tilted orientation.

\begin{figure}
\rotatebox{-90}{\includegraphics[height=8.5cm]{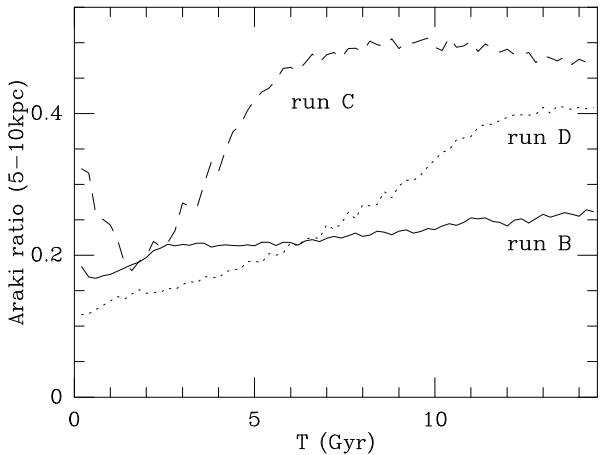}}
\caption{ Evolution for all runs of the $\sigma_z/\sigma_r$ ratio
averaged for radii between 5 and 10 kpc (weighted by the number of particles).}
\label{araki}
\end{figure}

Bending instabilities in galactic discs are expected when 
the ratio between $z$ and $r$ velocity dispersion 
($\sigma_z/\sigma_r$) is lower than the Araki (1985)'s value
of 0.3 (e.g. Revaz \& Pfenniger 2004). In our runs, this ratio is 
initially about 0.5 in the center, and falls towards 0.1 in the outer disc.
 Instabilities are therefore expected. At the end of the simulation, the 
$\sigma_z/\sigma_r$ ratio increases to be just above 0.3
(cf Figure \ref{araki}).

\begin{figure}
\rotatebox{-90}{\includegraphics[height=8.5cm]{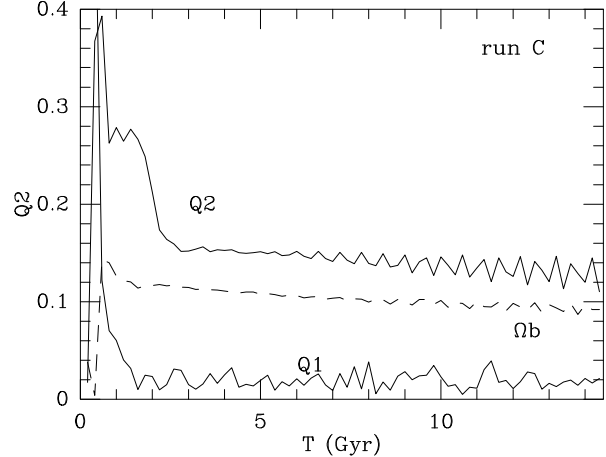}}
\caption{ Bar strength ($Q_2$) and bar pattern speed $\Omega_b$, for
run C. Units of $\Omega_b$ are 100km/s/kpc. Also shown is the 
$m=1$ strength $Q_1$.}
\label{bar-runC}
\end{figure}

In the edge-on views of Fig. \ref{cont-runC},  it is easy to
see the shape of the tilt (see  the T= 9.6 Gyr epoch for instance).
The center of the disc until a radius of 11kpc is tilted
with a position angle larger than 90$^\circ$, while the outer parts
are tilted the opposite way (with position angle smaller than 90$^\circ$).
The disc shows a warp, with a line of nodes following the peak of the $m=1$ 
perturbation. The line of nodes
is almost straight. The $A_1$ amplitude is also conspicuous 
inside and outside  a radius of about 14kpc, as is
visible in Fig. \ref{a1-toutr}. This is due to the off-centring
of the central part which is opposite of that of the outer parts.
  The limiting radius corresponds roughly to the resonant radius 
where $\Omega - \kappa$ = 0.

\begin{figure*}
\includegraphics[height=17cm,angle=-90]{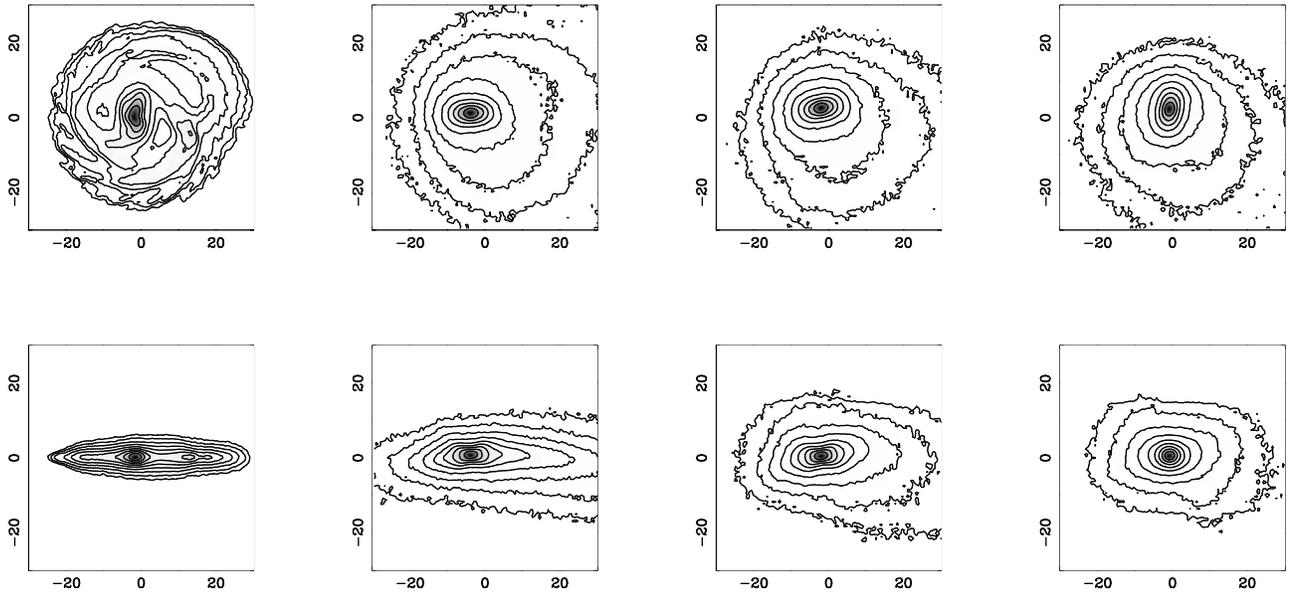}
\caption{ Run C: Contours in logarithmic scale of the surface density of
the stellar disc, face-on (top) and edge-on (bottom), at four different
epochs: T=0.8, 5.6, 9.6, 14.4 Gyr, from left to right. The contours are taken between
the maximum surface density, and a minimum equal to 5 10$^{-4}$ that value. In units of
900 M$_\odot$/pc$^2$, in square pixels of 470pc in size, the maxima are respectively
1.2, 1.7, 1.8, 1.9 for the face-on values, and 3.6, 2.5, 2.5 and 3.2 for the edge-on values.}
\label{cont-runC}
\end{figure*}

\begin{figure}
\rotatebox{-90}{\includegraphics[height=8.5cm]{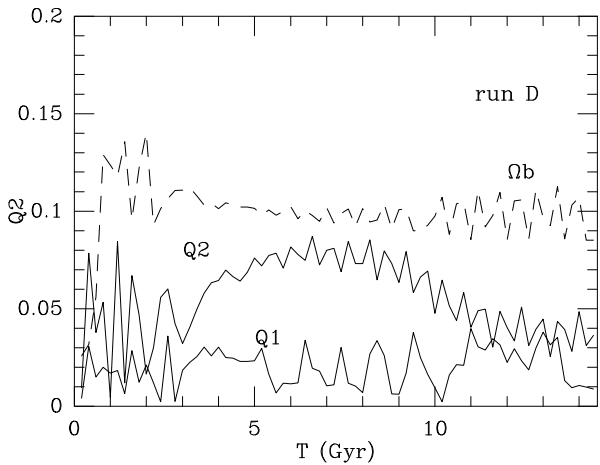}}
\caption{ Bar strength ($Q_2$) and bar pattern speed $\Omega_b$, for
run D. Units of $\Omega_b$ are 100km/s/kpc. Also shown is the 
$m=1$ strength $Q_1$.}
\label{bar-runD}
\end{figure}

\begin{figure*}
\includegraphics[height=17cm,angle=-90]{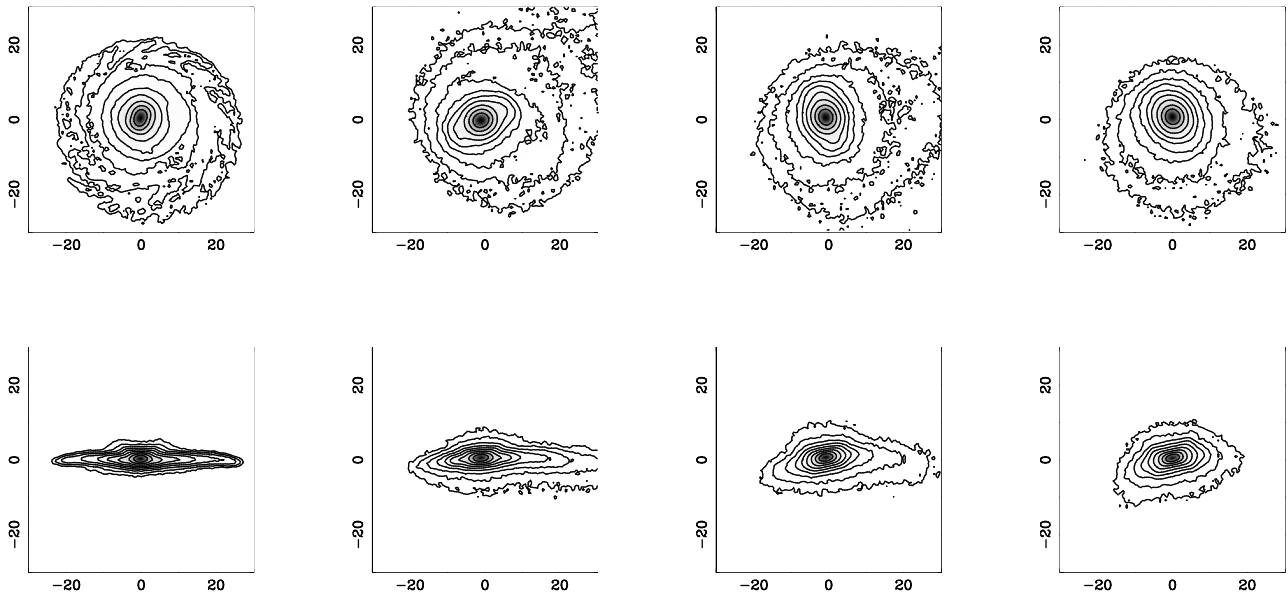}
\caption{ Run D: Contours in logarithmic scale of the surface density of
the stellar disc, face-on (top) and edge-on (bottom), at four different
epochs: T=0.8, 5.6, 9.6, 14.4 Gyr, from left to right. The contours are taken between
the maximum surface density, and a minimum equal to 5 10$^{-4}$ that value. In units of
900 M$_\odot$/pc$^2$, in square pixels of 470pc in size, the maxima are respectively
1.7, 1.7, 1.7, 1.7 for the face-on values, and 4.4, 3.8, 4.0 and 3.9 for the edge-on values.}
\label{cont-runD}
\end{figure*}

Run D was carried out with a twice smaller exponential scale-length, in order to explore the varying smoothing length parameter.  The latter is equal to the cell size of 234 pc, so that the ratio of the disc characteristic size to the softening length is reduced by a factor 2. The resulting disc shows a larger instability than for the reference run B, as can can be seen in the evolution of pattern strength in Fig \ref{bar-runD}, and in the contour plots of Figure \ref{cont-runD}. 

At first, it may appear paradoxical that run D, characterized by a larger softening with respect to the disc scale, is more unstable to the global $m=1$ mode.  For usual WKB local waves, self-gravity is reduced by the softening, and disc should be more stable. But for these global modes, involving the whole disc, with a wavelength comparable to the disc size, this is different. The instabilities are due only to the collective effect of gravity, and the softening mimicks the effect of pressure, which stabilises more local modes. The disc is then colder, and can sustain global modes.

\begin{figure}
\rotatebox{-90}{\includegraphics[height=8.5cm]{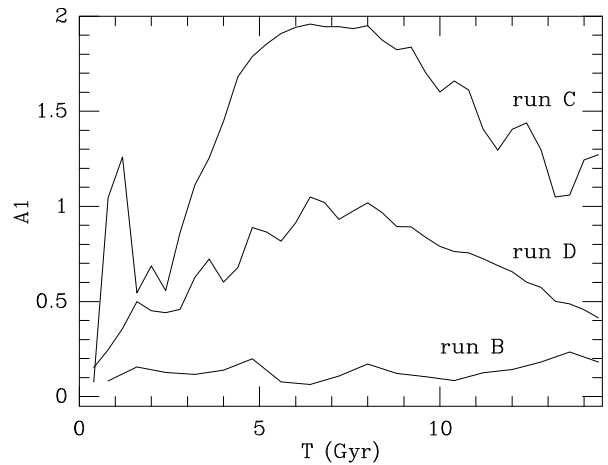}}
\caption{ Strength  of the $m=1$ mode, for
run B, C and D.  $A_1$ is the $m=1$ Fourier component of the
surface density, normalised with the $m=0$ component,
and taken at R=5kpc. }
\label{m1-toutr}
\end{figure}

The time evolution of the $m=1$ mode is
compared for all runs in Fig \ref{m1-toutr},
where is plotted the value of $A_1$ estimated at the radius of 5kpc.
A maximum at a value of $A_1 \sim 2$ is reached around
T= 7 Gyr for run C, and at about the same time for run D,
at $A_1 \sim 1$. For run B  the value is
negligible and tends to be stationary at $\sim 0.1$.
This evolution corresponds to a coherent lopsidedness deformation of 
the stellar disc.  The deformation appears quasi still in phase,
and within the noise it is impossible to distinguish a value
for the pattern speed $\Omega_1$ different from zero.

  The $m=1$ perturbations found in the present simulations
might appear quite large, with respect to what is usually found in
galactic simulations. However, this is due to the fact that in general
galaxy models are assumed to be embedded in massive spherical haloes,
which stabilise the discs. Here we have purely stellar discs, initially
without any central bulge component. Simulations with massive discs have also
been run by  Revaz \& Pfenniger (2004), who also find $m=1$
perturbations quite common.

\section{Comparison between analytical analysis and numerical simulations}
 
  The numerical simulations of an isolated purely 
stellar exponential disc reveals some confirmation of the analytical
derivations about the global mode. Although the $m=2$ perturbation
is always very strong, which is a feature not considered in the calculations,
it is possible to see the development of an $m=1$ perturbation, which 
appears to involve the whole disc, with a global characteristic. The wavelength
is long, at least of the same size as the disc.

The instability in the simulations appear for values
of the Toomre parameter larger than is predicted 
for the analytical calculations.  But the latter are computed for
an infinitely thin disc, and do not take into account all physical
parameters, except with approximations. The 
"effective" Q in the analytical model should be larger than computed, 
to take  account of the pressure, the softening, the thickness of the plane,
for instance.  
Merritt \& Sellwood (1994) found also that large-scale modes
could develop in numerical galaxies, well beyond the predictions of 
the analytical calculations in a thin sheet.

\begin{figure}
\rotatebox{-90}{\includegraphics[height=8.5cm]{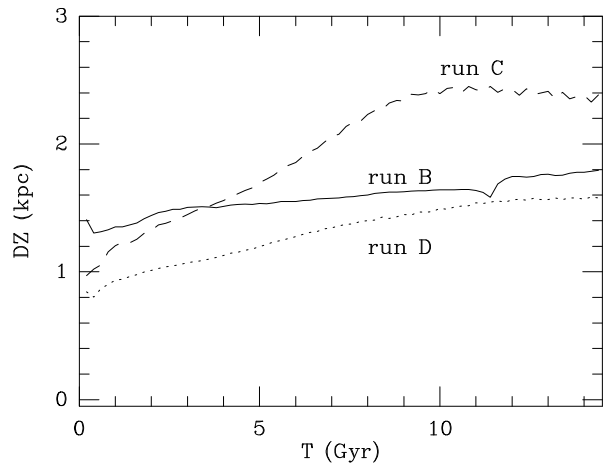}}
\caption{ Average thickness of the plane (variance of the z-coordinate
averaged over all particles) for 
run B, C and D.  The time evolution reveals a saturation 
after 10 Gyr.}
\label{dz-toutr}
\end{figure}

All runs develop a bar, a strong $m=2$ perturbation, that slowly develops
a peanut instability. The disc is buckling and thickening around the inner
Lindblad resonances, between 4 and 6 kpc in radius for runs B and C (3kpc for run D).
  The disc thickening is also enhanced by the heating of the disc
through the gravitational instabilities, and saturates after 10 Gyr,
as shown in Fig. \ref{dz-toutr}.
  The thickening of the plane is not only enhanced by the peanut alone (as in run B), but 
also by its coupling with the $m=1$, leading to the plane tilt: the final thickness
is higher in run C where both are present (Fig. \ref{dz-toutr}).

\begin{figure}
\rotatebox{-90}{\includegraphics[height=8.5cm]{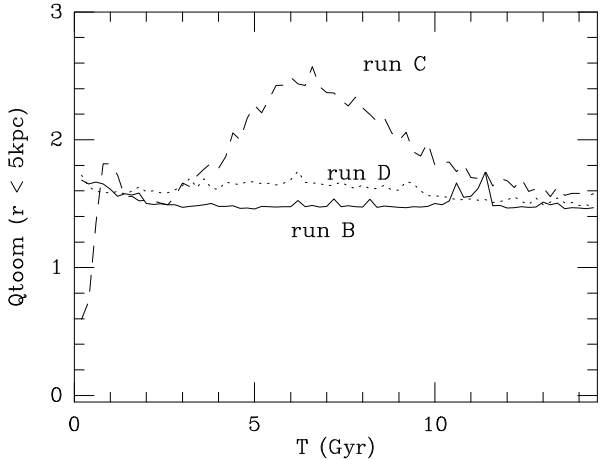}}
\caption{ Evolution of the Toomre parameter, for all runs,
averaged over the center of the disc, for radii lower than 5 kpc.
The average is weighted by the number of particles at each radius.}
\label{qtoom-r5}
\end{figure}

\begin{figure}
\rotatebox{-90}{\includegraphics[height=8.5cm]{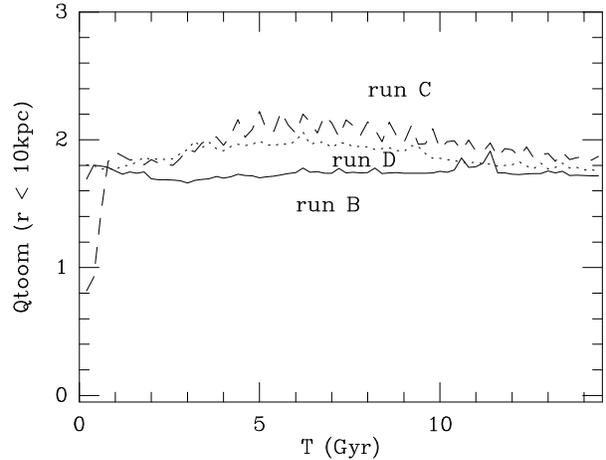}}
\caption{ Evolution of the Toomre parameter, 
averaged for radii lower than 10 kpc.
The saturation around $Q=2$ is then visible for all runs.}
\label{qtoom-r10}
\end{figure}

For a given softening length,
the $m=1$ develops essentially for the initially colder discs,
i.e. initial low values of the Toomre
parameter $Q$. The run B develops mainly a bar, with
little asymmetries, practically no $m=1$,
while the initially cold disc of run C develops
a pronounced lopsidedness.
 This strong instability heats violently the disc, which is
much hotter in the center for run C, as shown by the 
Toomre parameter averaged over the radii inside 5 kpc
(Fig. \ref{qtoom-r5}).  Since the disc is lopsided, the 
exact calculation of this parameter is not possible, 
the center of mass of the galaxy being different from
the maximum of density, or the minimum of the 
gravitational potential. However,
averaged over a much larger radius, the heating of the 
all runs become more similar (Fig. \ref{qtoom-r10}).

The pattern speed of the $m=1$ is very slow, such that it is difficult
to measure. The perturbation is almost still, difficult to estimate as prograde or retrograde, but certainly is partly retrograde at large radii. This confirms our semi-analytical results about the pattern speed of the lopsided mode.  

 A more interesting feature is the behaviour with respect to the relative
softening length $b$ (ratio of the softening to the exponential disc scale).
The instability with respect to the $m=1$ global
mode is larger for  a larger $b$ (run D), than for a smaller  $b$ (run B),
at the same value of $Q$.  This can be interpreted in terms of global mode
due only to collective gravity effects.

This behaviour might appear surprising, as the softening is known
to stabilize discs with respect to gravitational instabilities.
However, the softening suppresses essentially the small scale
perturbations, the modes with small wavelength (obeying the WKB
approximation).  This then leaves a disc more receptive to global modes,
and instabilities with wavelength of the order of the disc radius.

More quantitatively, Romeo (1994) has shown that the softening $s$
weakens the potential perturbations of radial wavenumber 
$k = 2 \pi / \lambda$, by a factor $ exp (- |k| s)$.  This efficiently
stabilises all small scale instabilities for run D, which is then 
colder in the first Gyr of evolution  and more able to develop
global modes.

Another argument that could explain that large wavelength
modes are favored in runD with respect to run B, is the 
order of magnitude of the self-gravitating critical scale length,
$\lambda_{crit} =  4 \pi^2 G \Sigma /\kappa^2 $ (cf Toomre 1981). 
Since the scale length $r_d$ of the disc in run D is 
divided by 2, keeping the total mass the same,
the surface density is multiplied by 4, and
$\kappa^2 $ is multiplied by 8. The critical scale
$\lambda_{crit}$ is divided by 2.  However, all
initial discs were truncated at R=25kpc, and the disc extent
is larger with respect to the critical scale in run D. 

\begin{figure}
\rotatebox{-90}{\includegraphics[height=8.5cm]{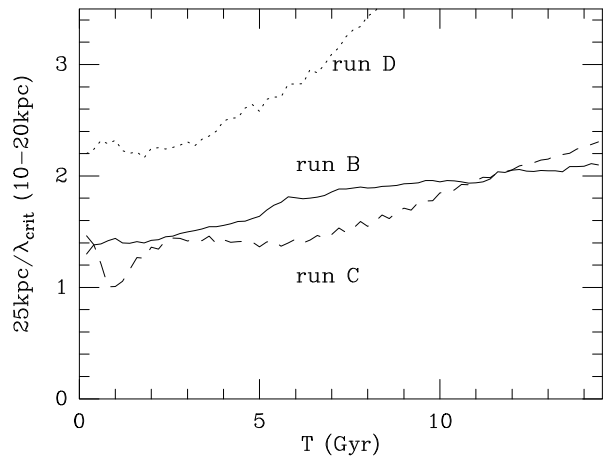}}
\caption{ Evolution of the $X$ parameter, 
indicator of the swing amplification efficiency,
averaged for radii between 10 and 20 kpc. }
\label{lamb20}
\end{figure}

The efficiency of the swing amplification is parametrized
by the $X$ factor (Toomre 1981)
where $X = \lambda/(\lambda_{crit} {\rm sin} i)$, and
$i$ is the pitch angle, nearly 90$^\circ$ here. For
wavelengths of the order of the disc extent, this $X$ factor
is initially larger for run D, as shown in Fig \ref{lamb20},
and approaches the optimum value for maximum amplification,
which is 2.

As in the calculations, the simulations show that the lopsided perturbation
can live several Gyrs, and even remain strong over a Hubble time.
This can solve the problem of maintenance of this perturbation in some
isolated galaxies.

\section{Conclusions}

By treating the lopsided distribution in an exponential galactic disc as a
global feature, and using the softened self-gravity of the perturbation, we show that the resulting lopsided mode is long lived to several Gyr (essentially increasing the winding up time scale), and it also exhibits a radially increasing amplitude as observed. {\it {The emerging pattern speed of the lopsided mode is extremely slow being a factor of $\sim$ 10 smaller than the local free precession rate and with no particularly preferred sense of precession}}. The behaviour of the pattern speed is confirmed in the numerical simulation.  
  
Our semi-analytical global treatment shows that {\it{the lopsidedness appears to be a purely oscillatory normal mode outside the continuum}} in smaller sized disc, while it emerges as an instability in comparatively larger sized disc with a very slowly precessing pattern. The e-folding growth time scale for a typical disc like that of Milky way is $\sim 1.2$ Gyr. 

Our numerical analysis show that the disc is never unstable to the lopsided modes without the indirect term arising due to the lopsided perturbation itself. Even though there is a subtle interplay between the indirect term due to the lopsided perturbation and the  collective effect of the softened self-gravity, it appears that the indirect term plays a crucial role in preparing the disc susceptible to lopsided instability.           

Numerical simulations of an isolated purely stellar component, an exponential disc without any stellar bulge or dark matter halo, has confirmed the theoretical predictions, and also show long-lived $m=1$ global modes. The criteria for stability are different than for local WKB modes,  and gravity softening stabilises essentially the local modes, but the global ones are not much affected by the softening. The $m=1$ modes superpose to the bar mode, that appear in all runs studied here. The buckling instability forming the peanut shape combines to the lopsidedness to produce disc tilts. 

Further work will include spheroidal collisionless components,
and also the gas component in the disc, in order to compare more realistically these results
with the observations.

\section*{Acknowledgments}
We thank the anonymous referee for a very useful and valuable comments which has improved our manuscript substantially. The numerical package LAPACK (see www.netlib.org) was used for solving the matrix 
eigenvalue problem. 
The 3-D computations have been realized on the Fujitsu
NEC-SX5 of the CNRS computing center, at IDRIS. 
K.S. would like to thank the CSIR-UGC, India for a Senior research fellowship. 
The work in this paper has been done in great part during
visits supported by the Indo-French centre (IFCPAR grant 2704-1).


\begin{thebibliography}{}
\bibitem[1989]{Adams89}  Adams, F. C., Ruden, S. P., \& Shu, F. H. 1989, ApJ, 347, 959

\bibitem[2006]{Angiras06}Angiras, R.A., Jog, C.J., Omar, A., \& Dwarakanath, K. S. 2006, MNRAS, 369, 1849 

\bibitem[1985]{Araki85} Araki S.: 1985, PhD Thesis, Massachusetts Institute of Technology

\bibitem[1980]{Baldwin80}Baldwin, J. E., Lynden-Bell, D., \& Sancisi, R. 1980, MNRAS, 193, 313 (BLS)

\bibitem[1987]{Ben87}Bienaym\'e O., Robin A., Cr\'ez\'e M.: 1987, A\&A 180, 94

\bibitem[1987]{Binney87} Binney, J. \& Tremaine, S. 1987, Galactic Dynamics (Princeton: Princeton Univ. Press)

\bibitem[1994]{Block94}Block, D. L., Bertin, G., Stockton, A., Grosbol, P., Moorwood, A. F. M., \& Peletier, R. F. 1994, A \& A, 288, 365

\bibitem[2005]{Bournaud05}Bournaud, F., Combes, F., Jog, C.J., \& Puerari, I. 2005, A \& A, 438, 507

\bibitem[1990]{Combes90}Combes F., Debbasch F., Friedli D., Pfenniger D.: 1990, A\&A 233, 82

\bibitem[1996]{Earn96} Earn, D. J. D. \& Lynden-Bell, D. 1996, MNRAS, 278, 395 

\bibitem[1998]{Evans98} Evans, N. W. \& Reed, J. C. A. 1998, MNRAS, 300, 106

\bibitem[1970]{Freeman70} Freeman, K. C. 1970, ApJ, 160, 811

\bibitem[1998]{Haynes98}  Haynes, M. P., van Zee, L., Hogg, D. E., Roberts, M. S., \& Maddalena, R. J. 1998, AJ, 115, 62 

\bibitem[1989]{Hozumi89}  Hozumi, S., \& Fujiwara, T. 1989, PASJ, 41, 841

\bibitem[2002]{Ideta02}  Ideta, M. 2002, ApJ, 568, 190

\bibitem[2001]{Jacobs01}  Jacobs, V., \& Sellwood, J. A. 2001, ApJ, 555, L25

\bibitem[1977]{James77}James, R.~A., 1977, J. Comput. Phys. 25, 71

\bibitem[1997]{Jog97}  Jog, C. J. 1997, ApJ, 448, 642 

\bibitem[1999]{Jog99}  Jog, C. J. 1999, ApJ, 522, 661

\bibitem[2001]{Kan01}  Kannappan, S. J., \& Fabricant, D. G. 2001, AJ, 121, 140

\bibitem[1991]{Kuijken91}Kuijken, K., Gilmore G.: 1991, ApJ 367, L9

\bibitem[1996]{Kuijken96}  Kuijken, K., Fisher, D., \& Merrifield, M. R. 1996, MNRAS, 283, 543


\bibitem[1998]{Levine98}  Levine, S. E., \& Sparke, L. S. 1998, ApJ, 496, L13

\bibitem[1989]{Lewis89} Lewis, J. R. \& Freeman, K. C. 1989, AJ, 97, 139

\bibitem[1999]{Lov99}  Lovelace, R. V. E., Zhang, L., Kornreich, D. A., \& Haynes, M. P. 1999, ApJ, 524, 634 

\bibitem[1965]{Bell65} Lynden-Bell, D. 1965, MNRAS, 129, 299    

\bibitem[2006]{Martin06} Martinez-Valpuesta, I., Shlosman, I., Heller, C. 2006, ApJ 637, 214

\bibitem[1990]{Mathur90} Mathur, S. D. 1990, MNRAS, 243, 529

\bibitem[1994]{Merritt94} Merritt D., Sellwood J.A.: 1994, ApJ 425, 551

\bibitem[2002]{Papaloizou02}Papaloizou, J.C.B., 2002, A\&A, 388, 615

\bibitem[1991]{Pfenniger91}Pfenniger D., Friedli D. 1991, A\&A ,252, 75

\bibitem[2004]{Polyachenko04} Polyachenko, E. V. 2004, MNRAS, 348, 345

\bibitem[2004]{Revaz04} Revaz, Y., Pfenniger D.: 2004 A\&A  425, 67

\bibitem[1994]{Richter94}  Richter, O. -G., \& Sancisi, R. 1994, A \& A, 290, L9 

\bibitem[1995]{Rix95}  Rix, H.-W., \& Zaritsky, D. 1995, ApJ, 447, 82

\bibitem[1994]{Romeo94}  Romeo, A.B.  1994, A\&A  286, 799


\bibitem[2006]{Saha06}  Saha, K., \& Jog, C. J. 2006, MNRAS, 367, 1297 



\bibitem[1997]{Sell97}  Sellwood, J. A., \& Valluri, M. 1997, MNRAS, 287, 124

\bibitem[1999]{Statler99}  Statler, T. S. 1999, ApJ, 524, L87

\bibitem[1964]{Toomre64}  Toomre, A. 1964, ApJ, 139, 1217

\bibitem[1981]{Toomre81}  Toomre, A. 1981, in ``The structure and evolution of normal galaxies'',
ed. S.M. Fall \& D. Lynden-Bell, Cambridge Univ. Press.

\bibitem[2001]{Tremaine01}  Tremaine, S. 2001, AJ, 121, 1776
\bibitem[2004]{Wilcots04} Wilcots, E.M., \& Prescott, M.K.M. 2004, AJ, 127, 1900

\bibitem[1997]{Zar97}  Zaritsky, D., \& Rix, H. -W. 1997, ApJ, 477, 118

\end{thebibliography}
\end{document}